\documentclass[sigconf, authorversion]{acmart}

\AtBeginDocument{%
  }

\usepackage{booktabs}
\usepackage{makecell}
\usepackage{graphicx}
\usepackage{subcaption} 
\usepackage{hyperref}

\begin{document}

\title[Gen-AI Personalization and VR Immersion in Oral Storytelling]{Wrapped in Anansi's Web: Unweaving the Impacts of Generative-AI Personalization and VR Immersion in Oral Storytelling}

\author{Ka Hei Carrie Lau}
\affiliation{%
  \institution{Technical University of Munich}
  \city{Munich}
  \country{Germany}}
\email{carrie.lau@tum.de}

\author{Bhada Yun}
\affiliation{%
  \institution{University of California, Berkeley}
  \city{Berkeley}
  \country{USA}}
\email{bhadayun@berkeley.edu}

\author{Samuel Saruba}
\affiliation{%
  \institution{Technical University of Munich}
  \city{Munich}
  \country{Germany}}
\email{ge36sev@mytum.de}

\author{Efe Bozkir}
\affiliation{%
  \institution{Technical University of Munich}
  \city{Munich}
  \country{Germany}}
\email{efe.bozkir@tum.de}

\author{Enkelejda Kasneci}
\affiliation{%
  \institution{Technical University of Munich}
  \city{Munich}
  \country{Germany}}
\email{enkelejda.kasneci@tum.de}

\renewcommand{\shortauthors}{Lau et al.}

\begin{abstract}
Oral traditions, vital to cultural identity, are losing relevance among youth due to the dominance of modern media. This study addresses the revitalization of these traditions by reconnecting young people with folklore. We introduce \textbf{Anansi the Spider VR}, a novel virtual space that combines first-person virtual reality (VR) with generative artificial intelligence (Gen-AI)-driven narrative personalization. This space immerses users in the Anansi Spider story, empowering them to influence the narrative as they envision themselves as the ``protagonists,'' thereby enhancing personal reflection. In a $2\times2$ between-subjects study with 48 participants, we employed a mixed-method approach to measure user engagement and changes in interest, complemented by semi-structured interviews providing qualitative insights into personalization and immersion. Our results indicate that personalization in VR significantly boosts engagement and cultural learning interest. We recommend that future studies using VR and Gen-AI to revitalize oral storytelling prioritize respecting cultural integrity and honoring original storytellers and communities.
\end{abstract}




\maketitle

\section{Introduction}

\begin{quote}
    ``Folklore at its best addresses both tradition and innovation and shows how constancy and change are interlinked in the dynamic process of civilization.'' \\- Wolfgang Mieder, 1987~\cite{mieder2015tradition}
\end{quote}

Oral traditions are vital in fostering cultural identity and continuity, serving as vessels for transmitting traditional knowledge and values across generations. On a societal level, oral traditions preserve cultural heritage while evolving to address modern societal challenges, contributing to societal resilience and continuity~\cite{toelken1996dynamics, mieder2015tradition, DNoyes2012}. On a personal level, they reinforce our origins, identity, and sense of belonging~\cite{DNoyes2012}. Toelken underscores the role of oral traditions, specifically folklore, in forging community identity through shared narratives, essential for preserving languages, customs, and values not commonly captured in written records~\cite{toelken1996dynamics}.

UNESCO’s heritage framework seeks to identify, preserve, and promote cultural heritage worldwide, particularly intangible cultural heritage (ICH) at risk of disappearing~\cite{UNESCO_BasicText_2003}. While this framework is essential for preventing the commodification and loss of authentic traditions, it can sometimes oversimplify complex cultural narratives~\cite{Hafstein2012}. These traditions continue to face challenges, including the lingering effects of colonialism~\cite{kotut2024griot}, the transformative impact of modern media~\cite{toelken1996dynamics}, and the pressures of globalization~\cite{Lubis2023}. Moreover, diminishing interest from younger generations~\cite{Putra2023, Cintya_2023} underscores the need for proactive preservation strategies to safeguard this invaluable cultural heritage.

While UNESCO’s efforts to preserve ICH are crucial, oral traditions, in particular, face unique challenges. Often passed down through spoken words and rarely formally documented, these traditions are fluid and vulnerable to loss. Previous efforts to preserve and revitalize these folklores, such as ``Digital Folklore''~\cite{Melnikova2020} and ``Folk-Computing''~\cite{borovoy2001folk}, have attempted to revolutionize folklore transmission. However, these approaches were limited by passive engagement and a lack of immersive experience, constrained by the technology available at the time.

With the advent of Generative Artificial Intelligence (Gen-AI) and Virtual Reality (VR), there is potential for these technologies to create dynamic and engaging platforms for preserving and sustaining cultural narratives. Previous research has explored the potential of personalization to enhance cultural engagement~\cite{bdcc6030073, Ardissono2012}, and VR’s potential role in enriching presence and embodiment~\cite{Kenderdine2015, Bakhtiary2023}. In this regard, our research investigates the combined effects of Gen-AI personalization with VR immersion, not only to preserve but also to breathe new life into traditional stories. This approach allows participants to deeply engage with narratives, bridging cultural gaps while reflecting on embedded values. By integrating these technologies, we aim to safeguard endangered cultures and enhance our understanding of minority cultures. This aligns with research showing how digital storytelling fosters social inclusion by providing reachable cultural experiences~\cite{Giglitto2019, Deidre2012, Nisi2023, Bratitsis2022, Ashlee2013}.

In this research, we chose `Anansi the Spider' as the foundation for our VR experience due to its cultural significance and educational role within Akan traditions, where Anansi symbolizes kingship, storytelling, and power structures~\cite{MEKDakubu1990}. Anansi's tales, renowned for using wit and trickery to impart moral lessons and challenge societal norms, also resonate with Afro-Caribbean experiences during the transatlantic slave trade, serving both as a connection to ancestral heritage and as symbols of resistance against colonial oppression~\cite{kelley2007magic, marshall2019not}. The enduring nature of Anansi stories, which have fostered community and subversive survival, underscores how folklore can facilitate intercultural dialogue and address modern challenges, as emphasized in~\cite{abba2016transcending}. By leveraging personalized Gen-AI for self-reflection and moral awareness, along with VR for active involvement in storytelling, these emerging technologies provide opportunities to engage users in cultural reflection and ethical decision-making.

To this end, a $2\times2$ between-subjects design was conducted with 48 participants to evaluate the synergies between Gen-AI and VR, particularly regarding the use of mediums (VR vs. Non-VR) and personalized content for folklore transmission (Personalized vs. Non-Personalized). Preliminary findings revealed that VR significantly enhances engagement and interest in cultural learning. Additionally, Gen-AI-driven personalization amplifies VR immersion, promoting self-reflection. Semi-structured interviews showed that participants in the personalized VR condition demonstrated not only a stronger connection to the story but also deeper levels of reflection. This suggests that personalized and immersive storytelling offers a promising approach to preserving oral traditions in modern contexts. To this end, the contributions of this paper are fourfold and as follows:

\begin{itemize} 
    \item We designed a VR experience that captures the essence of Anansi the Spider, incorporating interactive elements to engage contemporary audiences.
    \item We developed a narrative framework that integrates traditional storytelling with interactive, personalized user experiences.
    \item Our findings indicate that VR immersion significantly enhances engagement, while Gen-AI-driven personalization deepens immersion, fostering greater self-reflection and connection to the story.
    \item We constructed a thematic map (key themes: Immersion, Personalization, Engagement, Reflection, and Cultural Relevance) that delineates how VR and Gen-AI independently and complementarily contribute to storytelling experiences. Findings indicate that VR provides sensory and emotional immersion, essential for engaging narratives, while Gen-AI enhances user involvement and cultural connections through personalization.
\end{itemize}

\section{Background and Related Work}
The rapid advancement of Gen-AI technologies has led to their increasing application across various domains, including cultural heritage preservation and storytelling. Our study investigates the opportunities and limitations of Gen-AI in conjunction with VR for enhancing the transmission of oral traditions. This section provides an overview of research at the intersection of Gen-AI, VR, and cultural heritage preservation.

\subsection{Oral Traditions and Intangible Cultural Heritage}
UNESCO defines ICH as living expressions passed down through generations, encompassing oral traditions, performing arts, and social practices \cite{unesco2024, unesco2024oral}. These traditions are crucial in preserving cultural diversity and transmitting knowledge, values, and collective memory. The concept of ICH emphasizes that traditions are not static but evolve and adapt over time \cite{unesco2024, toelken1996dynamics, sims2005living}. Toelken's ``Dynamics of Folklore'' posits that folklore possesses both dynamic (changing) and conservative (static) features, allowing for adaptability while maintaining continuity \cite{toelken1996dynamics, sims2005living}. Mieder, on the other hand, argues how folklore can evolve creativity to remain relevant in modern context~\cite{mieder2015tradition}.

The fragility of oral traditions in the face of modernization has sparked efforts to safeguard them using various technological approaches. Borovoy et al.'s ``Folk Computing'' concept demonstrates early attempts to digitally augment oral traditions~\cite{borovoy2001folk}, while Lu et al. explored the use of live streaming to engage audiences with traditional cultural practices~\cite{lu2019stream}.

However, the digitization and adaptation of oral traditions raise important ethical considerations. Kotut et al.'s griot-style methodology emphasizes ethical collaboration and community involvement in preserving indigenous knowledge \cite{kotut2024griot}, arguing for culturally specific approaches rather than generalized design practices \cite{noe2024equitable, Guerrero_Millan}. These perspectives underscore the need for carefully designed and community-driven approaches in adapting traditional stories like Anansi folktales to new media formats, balancing innovation with cultural integrity and respect for the original storytellers and communities.

\subsection{Virtual Reality for Immersive Storytelling}
VR has emerged as a powerful tool for preserving and communicating ICH, offering immersive experiences that enhance user engagement compared to traditional methods~\cite{selmanovic2020accessibility, innocente2023framework, Maud2021}. Studies have shown VR’s potential to bring oral traditions to life, as demonstrated in projects such as the adaptation of African folktales~\cite{skovfoged2018tokoloshe}. However, maintaining cultural authenticity and achieving effective visual representation remain challenges. The multi-sensory capabilities of VR enable more interactive, engaging, and personalized experiences~\cite{li2024multisensory, Wu_CHI_2024}, particularly for users from diverse cultural backgrounds who may lack access to physical cultural sites or context~\cite{innocente2023framework, selmanovic2020accessibility}. These immersive features not only enhance user engagement but also play a crucial role in cultural preservation by allowing users to experience stories in more meaningful ways.

Beyond immersion, recent advancements in VR techniques, such as perspective-shifting~\cite{dollinger2024swapping} and accurate avatar representation~\cite{ponton2024misalignment}, further deepen user involvement. These features not only enhance believability but also foster empathy and cultural understanding through role-playing~\cite{rifat2024empathy}, aligning directly with the goal of using VR to preserve and share cultural heritage

In addition to immersion techniques, advances in narrative design have expanded the possibilities for immersive storytelling. The use of design cards to inspire augmented experiences~\cite{neuhaus2024unreality}, AI-based narration tools~\cite{Alvarez2022}, and the concept of ``story living'' to create immersive narratives~\cite{Vallance2022} all highlight the growing potential of this field. However, as VR environments become more immersive, ethical considerations, such as memory source confusion~\cite{bonnail2024realism}, must be addressed. When adapting culturally rich folktales like `Anansi', it is critical that these technologies maintain a clear distinction between virtual and real experiences, ensuring cultural preservation and narrative integrity.

Building on previous research, Bahng et al. identified four reflexive dimensions for designing VR experiences that encourage deep, personal reflection~\cite{bahng2020reflexive}. Drawing inspiration from this methodology, our Anansi project seeks to create VR folktale experiences that prompt reflection on embedded moral lessons and cultural values. To evaluate the effectiveness of our approach, we employ semi-structured interviews to capture the nuanced and personal nature of participants' experiences with the Anansi VR stories.

\subsection{Gen-AI-driven Narrative Personalization}
Gen-AI-driven narrative personalization has evolved significantly since early attempts like the Storytelling Agent Generation Environment (SAGE), which allowed children to create personalized interactive storytelling experiences~\cite{umaschi1996sage, Kucirkova2016}. Recent advancements, such as generative agents, have paved the way for dynamic, human-like responses that autonomously react to user or non-playable character (NPC) interactions and environmental changes~\cite{park2023simulacra, Bozkir_2024, MatyasTheEO}. These developments align closely with the goals of creating personalized and culturally relevant folktale experiences in projects like the Anansi storytelling system.

While the accuracy of Gen-AI remains a growing concern due to issues like hallucinations and factual errors~\cite{yao2024llmlieshallucinationsbugs}, the potential applications of Gen-AI in personalizing narratives extend beyond entertainment into educational and cultural domains. In project-based learning scenarios, Gen-AI has been envisioned to support divergent thinking~\cite{cropley2023artificial}, provide feedback~\cite{Kathrin2023}, and guide learning~\cite{zheng2024projects}. Similarly, in cultural heritage contexts, large language models have been employed to tailor virtual museum tours to individual preferences~\cite{wang2024multimodal}, enhancing narrative depth and personal relevance~\cite{Wu2024}. These applications demonstrate the versatility of Gen-AI in creating engaging, personalized experiences across various fields.

While Gen-AI offers vast potential, its effectiveness depends on whether the design takes cultural differences in user preferences and expectations into account. Prior research has shown that applying AI in Indigenous contexts poses challenges, particularly in the under-representation of cultural nuances~\cite{Guerrero_Millan}. Additionally, cultural models of self and environment shape people's desired interactions with Gen-AI~\cite{ge2024culture}. This highlights the importance of developing culturally conscious Gen-AI personalization strategies, especially when working with traditional narratives like the Anansi folktales. By incorporating these cultural nuances, Gen-AI-driven narrative personalization can create more meaningful and resonant experiences for diverse audiences, potentially enhancing engagement with and preserving ICH.

\subsection{Synergy of Gen-AI and XR in Interactive Experiences}

The intersection of Gen-AI and XR offers powerful potential for personalized and adaptive storytelling, merging dynamic content creation with immersive user engagement. Research by Hirzle et al.~\cite{hirzle2023scoping} and Bozkir et al.~\cite{Bozkir_2024} has explored the intersection of extended reality (XR) and AI, outlining key potentials: the creation of virtual environments, understanding users, and facilitating interactions with intelligent virtual agents. Our work builds on these insights by exploring how Gen-AI-driven narrative personalization in VR storytelling leverages these advances, deepening user engagement with culturally rich narratives.

Kenderdine et al.~\cite{Kenderdine2014} addresses the challenges of digitizing ICH, proposing the concept of `embodied museography,' which emphasizes immersion, interaction, and participation. This concept aligns with our objective of using VR to create interactive storytelling experiences that capture the evolving nature of oral traditions through immersive technologies.

The integration of Gen-AI and XR opens new frontiers in interactive and creative storytelling. For example, Tan et al.'s AudioXtend~\cite{tan2024storytelling} demonstrates how AI-generated visuals can enhance audiobook experiences, improving both content recall and narrative engagement. Similarly, Döllinger et al.~\cite{dollinger2024swapping} showcase how real-time perspective switching in VR enables dynamic, multi-perspective narratives. This approach aligns with our research goal of combining third-person narration with first-person experiences in Anansi storytelling, allowing users to engage more deeply with cultural narratives and reflect on personal and communal experiences.

Research in cultural spaces and heritage preservation demonstrates how personalized and adaptive VR experiences can increase accessibility to cultural content. De La Torre et al.'s LLMR framework exemplifies this by creating 3D mixed reality scenes tailored to users with varying levels of expertise, needs, and backgrounds~\cite{delatorre2024mixed}. Moreover, the fusion of Gen-AI and immersive technologies presents exciting opportunities for cultural engagement. Studies have shown how these technologies enable participants to actively co-create cultural expressions, allowing traditional knowledge to evolve within digital environments~\cite{Pataranutaporn2024, lau2024_llm, FuKexue2024}. This fusion highlights the transformative potential of Gen-AI and VR in not only preserving but also expanding cultural narratives, making them adaptable and relevant in the digital age.

These advancements highlight the potential of combining Gen-AI and XR as powerful tools for cultural preservation, participation, and innovation, which informed the methods used in this study to explore their application in these areas.

\section{Methods}
To investigate the impacts of Gen-AI personalization and VR immersion on the storytelling experience of Ghanaian Anansi folktales, we conducted a user study approved by the Institutional Review Board (IRB) of Technical University of Munich with our prototype, `Anansi the Spider VR,' features a Gen-AI-driven NPC, Onini, allowing participants to interactively shape the traditional tale. This study compares the effects of personalized versus non-personalized story versions delivered through VR or visual slides (sequential images) and audio on user engagement, reflection, and cultural understanding, thereby helping to determine how technological implementations can enhance storytelling and cultural appreciation in virtual environments.

Our research aims to answer the following research questions (RQs):
\begin{itemize}
    \item \textbf{RQ1}: How does Gen-AI personalization affect user engagement and cultural learning interest in interactive storytelling experiences?
    \item \textbf{RQ2}: How does VR immersion influence user engagement and cultural learning interest in immersive narrative environments?
    \item \textbf{RQ3}: How do personalization and immersion synergize to enhance the storytelling experience and facilitate the transmission of intangible cultural heritage?
\end{itemize}

The following subsections describe the participants, apparatus, procedures, user study design, measures, and analyses.

\subsection{Participants}

A total of 48 participants were recruited from a university student population, including 21\% high school diploma, 42\% bachelor's degree, 31\% master's degree, and 6\% doctoral students, for this study. The participants' ages ranged from 22 to 46 years, with a mean age of 26.75 \((SD = 4.66)\), consisting of 22 males and 26 females. A pre-assessment questionnaire was administered to gauge their prior experience with digital storytelling, VR, oral traditions, African culture, and their interest in exploring other cultures. Notably, 40\% of the participants regularly engaged in digital storytelling activities such as listening to audiobooks and podcasts. Additionally, 42\% had some previous exposure to VR technology, while 67\% were completely unfamiliar with African folklore. Moreover, 54\% expressed a strong interest in learning about other cultures, and 44\% were comfortable with technology that personalizes content based on user information. Eligibility for participation required individuals to be at least 18 years old, possess normal or corrected-to-normal vision, and have the ability to understand and communicate in English fluently. Participants with a history of severe motion sickness were not considered for the study. Each participant was compensated with a \texteuro10 voucher for their involvement.

Semi-structured interviews were conducted with 20 participants from the initial group of 48. The backgrounds of these participants are detailed in Table~\ref{tab:ParticipantDemographics}. Interviews continued until data saturation was achieved, a point at which no new insights were observed, aligning with the guidance of Longhurst~\cite{Longhurst2003}, Hennink and Kaiser~\cite{HENNINK2022114523}, and Adams~\cite{Adams_2015}. These authors highlight saturation as essential in qualitative research to ensure that the data collection comprehensively covers the phenomenon under study. The interviews focused on personal reflections, perceived immersion, and emotional impacts, complementing the findings from the quantitative data.

\begin{table}[t]
  \caption{Summary of Participant Demographics from the Semi-Structured Interviews.}
  \Description{This table presents demographic information for participants involved in the semi-structured interviews. It includes details about each participant's gender, age, cultural background, experience with virtual reality (VR), and occupation.}
  \label{tab:ParticipantDemographics}
  \centering
  \resizebox{\linewidth}{!}{%
  \begin{tabular}{lccccc}
    \toprule
    \textbf{Interviewee} & \textbf{Gender} & \textbf{Age} & \makecell{\textbf{Cultural Background}} & \makecell{\textbf{VR Experience}} & \textbf{Occupation} \\
    \midrule
    VP1  & F  & 26 & Indian         & Moderately familiar  & Accountant           \\
    VP2  & F  & 23 & Turkish        & Very familiar        & EdTech Researcher     \\
    VP3  & M  & 25 & Indian         & Slightly familiar    & Network Planner       \\
    VP4  & F  & 28 & Peruvian       & Slightly familiar    & Consultant            \\
    VP5  & F  & 29 & Uzbek          & Moderately familiar  & Manager               \\
    VP6  & M  & 24 & Chinese        & Very familiar        & VR Engineer           \\
    VP7  & F  & 25 & Romanian       & Slightly familiar    & Software Engineer     \\
    VP8  & F  & 23 & \makecell{Indian, \\ German}  & Slightly familiar & Sustainability Consultant \\
    VP9  & M  & 26 & \makecell{Brazilian, \\ German} & Slightly familiar    & Electrical Engineer   \\
    VP10 & M  & 26 & \makecell{Ghanaian, \\ German} & Moderately familiar  & Business Developer    \\
    VNP1 & M  & 22 & Russian        & Very familiar        & Entrepreneur          \\
    VNP2 & M  & 35 & Kenyan         & Very familiar        & Manager               \\
    VNP3 & F  & 25 & Turkish        & Moderately familiar  & Kindergarten Teacher  \\
    VNP4 & M  & 24 & \makecell{Vietnamese, \\ American} & Moderately familiar  & ML Engineer           \\
    VNP5 & M  & 27 & Indian         & Very familiar        & Aerospace Engineer    \\
    VNP6 & F  & 24 & Peruvian       & Moderately familiar  & Student               \\
    NVP1 & M  & 31 & Indian         & Moderately familiar  & Product Manager       \\
    NVP2 & F  & 25 & Uzbek          & Moderately familiar  & Student               \\
    NVNP1 & M & 28 & Indian         & Very familiar        & Manager               \\
    NVNP2 & F & 24 & \makecell{Italian, \\ German} & Slightly familiar    & Student               \\
    \bottomrule
  \end{tabular}%
  }
\end{table}

\subsection{Apparatus}
\begin{figure}[ht]
  \centering
  \begin{subfigure}[t]{0.48\columnwidth}
      \centering
      \includegraphics[width=\linewidth, keepaspectratio]{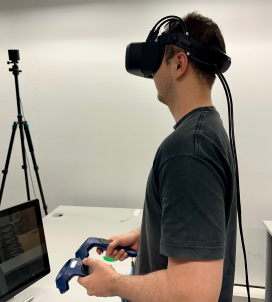}
      \caption{A participant using the Varjo VR-3 headset with HTC Vive controllers for the VR condition.}
      \label{fig:VRsetup}
  \end{subfigure}
  \hfill
  \begin{subfigure}[t]{0.48\columnwidth}
      \centering
      \includegraphics[width=\linewidth, keepaspectratio]{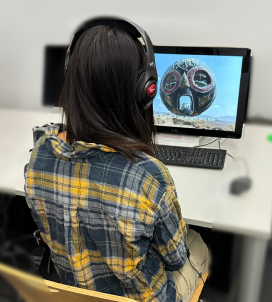}
      \caption{Non-VR setup featuring a Creative Fatal1ty Gaming Headset, Schenker i7-13700K RTX 4080 desktop, and a Dell P2314T monitor.}
      \label{fig:NonVRsetup}
  \end{subfigure}
  \caption{Experimental setup for VR and Non-VR conditions.}
  \Description{Figures show the experimental setup for VR and Non-VR conditions.}
  \label{fig:ExperimentSetup}
\end{figure}

The primary hardware platform used in this study was the Varjo VR-3 (Model HS-6). This head-mounted display (HMD) was powered by a GeForce RTX 4080 and an Intel Core i7-13700K processor with 32 GB of RAM, achieving an average frame rate of 90 frames per second. The hardware setup included HTC Vive Controller 2.0 and HTC Vive Steam VR Base Station 2.0, with VR interactions facilitated by Unity’s XR Interaction Toolkit. As depicted in Figure~\ref{fig:VRsetup}, the experimental setup shows a participant engaging with the VR environment. For the non-VR condition, as illustrated in Figure~\ref{fig:NonVRsetup}, the setup included a Schenker desktop with an i7-13700K RTX 4080 configuration, a Dell P2314T monitor, and a Creative Fatal1ty HS-800 Gaming Headset. The PC was powered by an Intel Core i7-13700K processor and an NVIDIA GeForce RTX 4080 graphics card, effectively supporting high-resolution displays. The monitor features a 23-inch (58.42 cm) display with a resolution of $1920 \times 1080$ pixels and a brightness of 300 cd/m². The headset covers a standard frequency range from 20Hz to 20kHz, enhancing overall immersiveness by delivering clear audio that complements the visual quality.

The VR game was developed using Unity 2021.3.33f1 LTS with the Universal Render Pipeline (URP) to optimize performance. The environment incorporated a mix of custom models and assets sourced from platforms such as Unity Asset Store, TurboSquid, CGTrader, and Sketchfab. Notable assets included the ``Tropical Forest Pack''\footnote{\url{https://assetstore.unity.com/packages/3d/environments/tropical-forest-pack-49391}, last accessed on 27 July 2024}, and character models such as Anansi's spider body\footnote{\url{https://www.cgtrader.com/free-3d-models/animals/insect/low-poly-spider-model}, last accessed on 27 July 2024}, Onini the Python\footnote{\url{https://sketchfab.com/3d-models/ball-python-74166573c34c40bc977c7a8db5b92ff4}, last accessed on 27 July 2024}, and Nyame, depicted by an Ashanti moon mask\footnote{\url{https://sketchfab.com/3d-models/ashanti-moon-mask-a9ea6430a5c44be998b94040712da159}, last accessed on 27 July 2024}. Custom assets such as Anansi's mask and clay huts were modelled in Blender specifically for this project.

For dialogue interaction, the integration of ChatGPT into Unity was achieved using an unofficial OpenAI Unity package \footnote{\url{https://github.com/srcnalt/OpenAI-Unity}, last accessed on 27 July 2024}. This package facilitated the use of OpenAI's API to generate dynamic dialogue and incorporate Whisper services for speech-to-text (STT) functionality. Text-to-speech (TTS) services were provided by Eleven Labs, an AI-powered voice synthesis tool, with API requests scripted package \footnote{\url{https://gist.github.com/THeK3nger/882a31f52bb002dac155ad95529c3680}, last accessed on 27 July 2024} on GitHub, enhancing the authenticity of the narrated content.

\subsection{Procedures}
Prior to participating in the study, eligible participants were selected based on pre-assessment criteria and invited to take part. Upon arrival, participants were briefed about the research goals and provided informed consent. They then underwent height calibration to ensure accurate tracking within the VR environment, which is crucial for tasks requiring precise spatial awareness \cite{Holzwarth2021, Wu_CHI_2024}. A five-minute tutorial followed, teaching them VR navigation, NPC interactions, and object manipulation techniques.

\begin{sloppypar}
The experiment began immediately after this preparatory phase. Following the VR experience, participants completed a post-assessment questionnaire to evaluate their engagement, story comprehension, and changes in cultural appreciation.
Semi-structured interviews were conducted with a select group of 20 participants (saturation point), chosen from four different experimental conditions.
\end{sloppypar}

Data collection was conducted in a controlled laboratory setting. Each session lasted approximately 50 minutes, including a 30-minute block for briefing, experimentation, and post-assessment, with an additional 20 minutes allocated for the semi-structured interviews.

\subsection{User Study Design}
Our user study utilized a \(2 \times 2\) between-subjects design, incorporating two independent variables: \textbf{personalization} (Personalized vs. Non-Personalized) and \textbf{medium} (VR vs. Non-VR). We assessed various dependent variables, including \textbf{user engagement} (measured by the User Engagement Scale), \textbf{story comprehension} (evaluated through the Story Comprehension Test), responses to the \textbf{cultural interest questionnaire}, and \textbf{conversation logs with NPCs}. Participants (\(n = 48\)) were randomly distributed into four groups (\(n = 12\) each): VR-Personalized (\textbf{VP}), VR-Non-Personalized (\textbf{VNP}), Non-VR-Personalized (\textbf{NVP}), and Non-VR Non-Personalized (\textbf{NVNP}), ensuring equal representation across conditions.

The personalized groups experienced dynamically adjusted narratives that incorporated personal details such as names and pronouns, while the non-personalized groups encountered fixed narratives. The Non-VR groups interacted with the story through visual slides and audio, in contrast to the immersive VR environment.

\begin{figure}[ht]
  \centering
  \includegraphics[width=\linewidth]{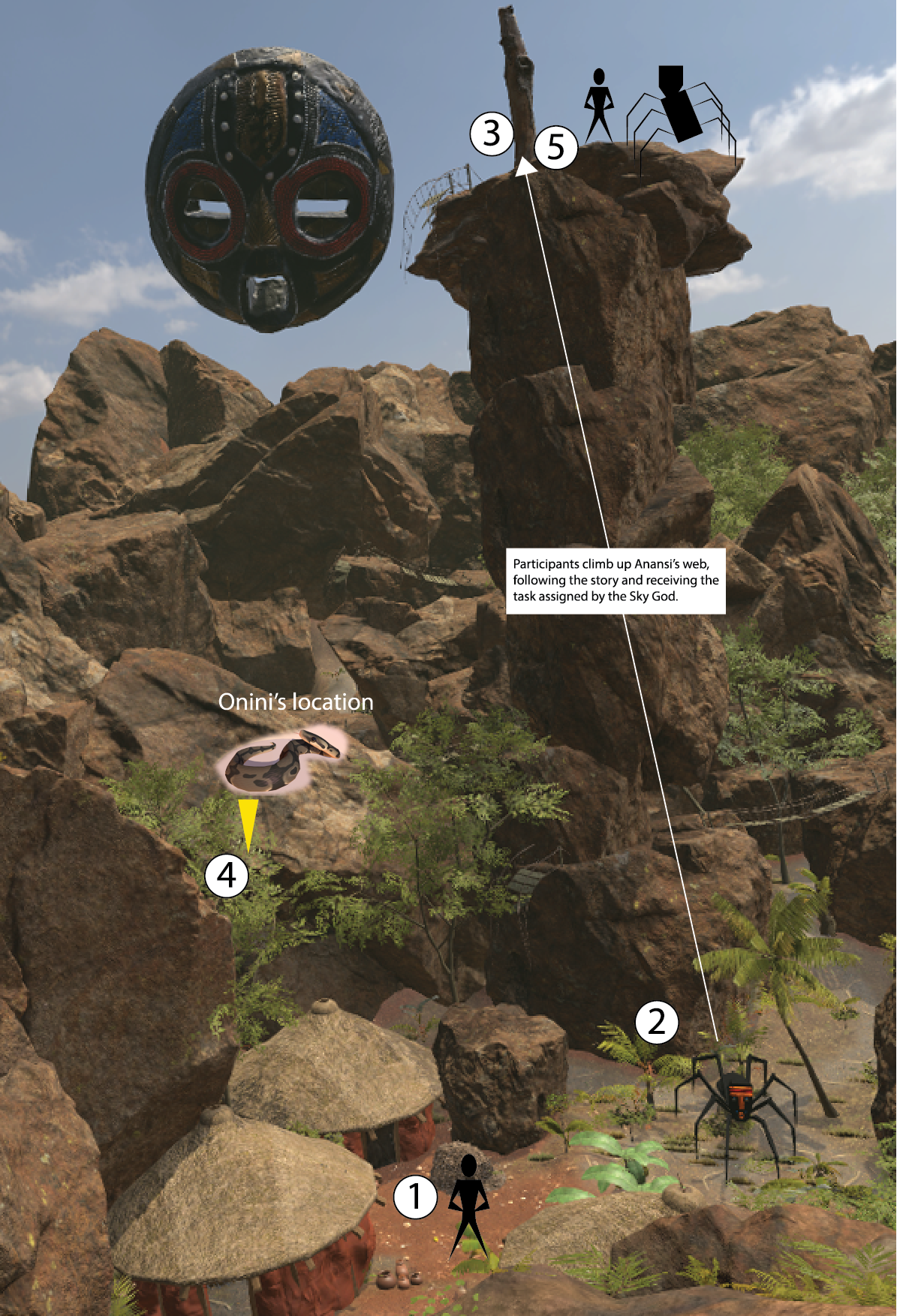}
  \caption{Interactive Narrative Path in `Anansi the Spider VR': Figure illustrates the five key stages of the VR narrative. Participants start their journey at Stage 1 in the village, walking along a jungle path to Stage 2 where they meet Anansi, climb Anansi's web at Stage 3 to receive tasks from the Sky God, search for Onini at Stage 4, and conclude the experience at Stage 5.}
  \label{fig:narrativePath}
  \Description{This figure illustrates the interactive narrative path within the 'Anansi the Spider VR' experience, highlighting five key stages. Participants begin their journey at Stage 1 in a village setting, proceed along a jungle path to reach Stage 2 where they encounter Anansi. At Stage 3, they climb Anansi's web to receive tasks from the Sky God, then move to Stage 4 where they search for Onini. The narrative concludes at Stage 5, completing the VR experience.}
\end{figure}

\begin{figure}[!htb]
  \centering
  \includegraphics[width=\linewidth]{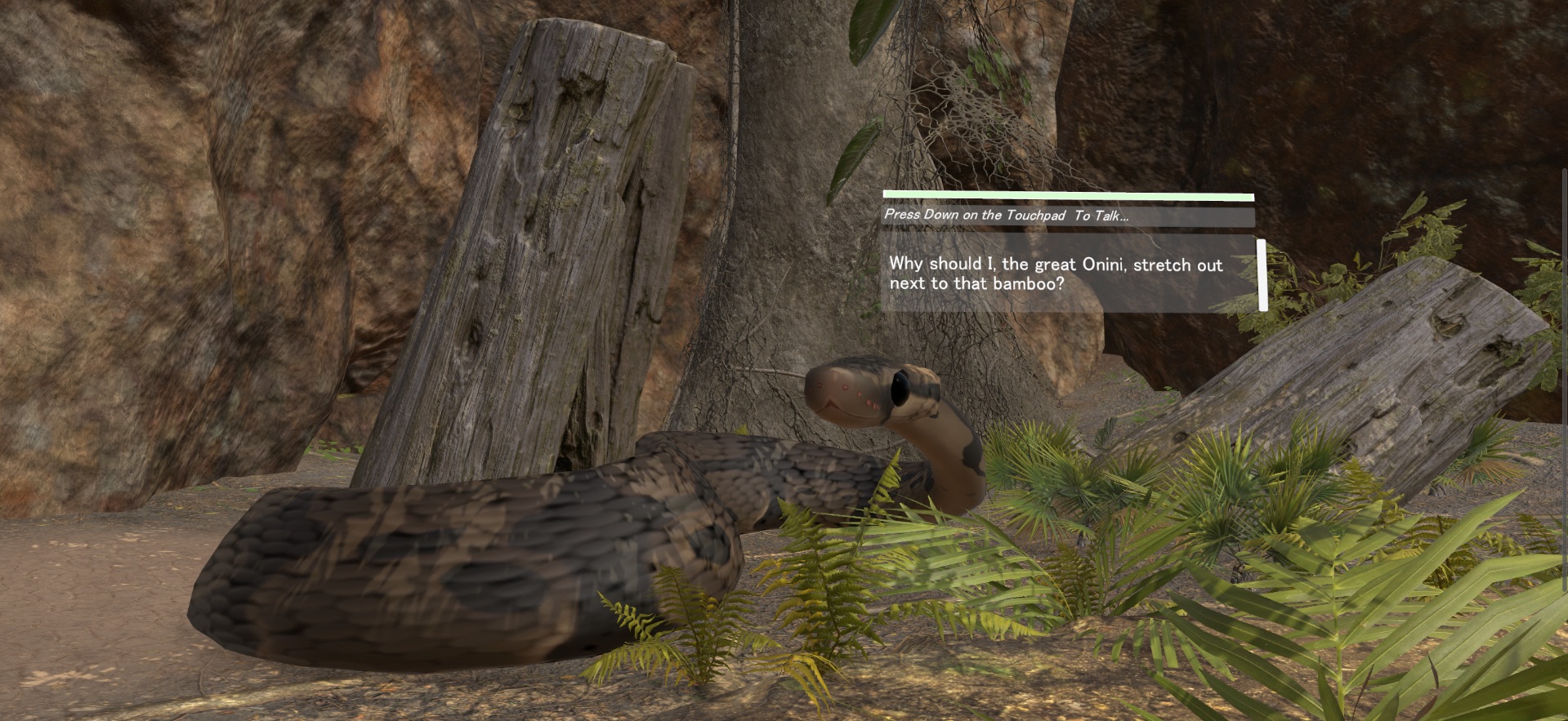} % Adjust width as necessary
  \caption{Interaction with Onini in the VR environment, where participants use Speech-to-Text options to persuade Onini to stretch alongside a bamboo stick, as part of their task in the `Anansi the Spider VR' narrative.}\label{fig:BaitOnini}
  \Description{This figure depicts the interaction with Onini in the 'Anansi the Spider VR' environment. Participants use Speech-to-Text options to persuade Onini to stretch alongside a bamboo stick as part of their task within the narrative. The interaction is a key element of the experience, requiring participants to engage verbally with the character to advance the story.}
\end{figure}

During the experiment, participants engaged in a structured sequence of activities, as detailed in Figure~\ref{fig:narrativePath}, which progressed through a narrative path beginning in a jungle setting. The key interactions included meeting Anansi, climbing a web, meeting the Sky God, and negotiating with the NPC Onini to stretch out for capture. The interaction with Onini, presented in Figure~\ref{fig:BaitOnini}, was facilitated by GPT-4, determining the success based on participants' vocal inputs. VR sessions were automatically logged in Unity, capturing details such as completion time, NPC interactions, and spatial coordinates. The non-VR sessions concluded with a video fade-out, signaling the end of the experience.

\subsection{Anansi's Spider Stories}

The background of this VR narrative experience is inspired by the Anansi tales, which hold a significant place in African folklore, particularly among the Akan people. In these stories, Anansi the Spider is depicted as a clever trickster and wise figure~\cite{marshall2007liminal, van2007anansi, pelton1989trickster}. These tales serve important educational purposes, conveying values such as intelligence, ingenuity, and community spirit through Anansi’s adventures. Furthermore, they preserve cultural heritage, functioning as a medium for passing down traditions and moral lessons across generations. This VR experience presents an adaptation of one of the most famous Anansi stories: \textit{How it came about that the Sky God’s stories came to be known as ``Spider-stories''}~\cite{rattray2023akan}. Together with the player, Anansi must capture Onini, the Python, to obtain all the stories in the world from the Sky God Nyame. The experience simplifies and streamlines the original story to fit the VR medium and the constraints of the study while preserving its core themes and moral teachings.

Central to this VR experience is the integration of Gen-AI, which enables natural, personalized interactions with NPCs, particularly with Onini, the Python. Gen-AI refers to systems capable of producing new content, such as text, speech, or images, based on the data they are trained on~\cite{Cohan2024}. In this case, the VR system leverages multiple Gen-AI models to bring the NPCs to life. OpenAI’s Whisper~\footnote{\url{https://openai.com/index/whisper/}, last accessed on 27 July 2024} is employed for speech-to-text (STT) conversion, transcribing the player’s spoken input into text. Then, ChatGPT, powered by GPT-4~\footnote{\url{https://openai.com/index/gpt-4/}, last accessed on 27 July 2024}, generates a contextual response as the NPC based on the player’s input. This ensures that the NPC’s reactions are personalized and aligned with the narrative. Eleven Labs~\footnote{\url{https://elevenlabs.io/}, last accessed on 27 July 2024} is used throughout the experience to dynamically generate the voices of the NPCs. This combination of technologies allows the NPCs to respond in real-time, creating an immersive and interactive player experience.

The prompt engineering process was designed to maintain narrative consistency while allowing for personalized player experiences. It involves several steps: First, we provide~\textbf{general instructions} to ChatGPT, which include contextual information about the game and ChatGPT's role. Secondly, we assign~\textbf{character attributes or personalities}, detailing specific traits of the NPC. Next, we provide ChatGPT with~\textbf{state descriptions}, offering a detailed context about the current game state. We then give~\textbf{scene context}, which involves descriptions of the surroundings and relevant NPCs. Lastly, we include~\textbf{script instructions} that consist of detailed directives on how NPCs should respond to the player's input and the specific information the NPC needs to convey.

For the character Onini, the Gen-AI technologies in our system (\textit{Whisper}, \textit{ChatGPT}, \textit{ElevenLabs}) work together to create a personalized and dynamic interaction between the player and NPC. In the game, the player must persuade Onini to stretch out fully next to a bamboo stick the player has placed beside her. ChatGPT, taking on the role of Onini, evaluates whether the player’s proposition is valid and persuasive enough and reacts in character. Eleven Labs is then used to generate the NPCs’ voices dynamically. Pre-recorded voices were used where dynamic generation was not necessary, such as for non-personalized conditions of the study.

The NPC dialogues, apart from Onini’s interactions and the inclusion of personalized player names, were pre-scripted. Custom voice profiles with West African accents were generated using Eleven Labs for the narrator and other NPCs, ensuring consistency and immersion throughout the experience. 

The game is structured into three primary scenes, each managed separately within Unity. Interaction within the game is linear, with an all-knowing narrator guiding the player and providing context at key moments. This structure ensures that the focus remains on the story, with minimal distractions from exploration.

\begin{figure}[ht]
  \centering
  \includegraphics[width=\linewidth]{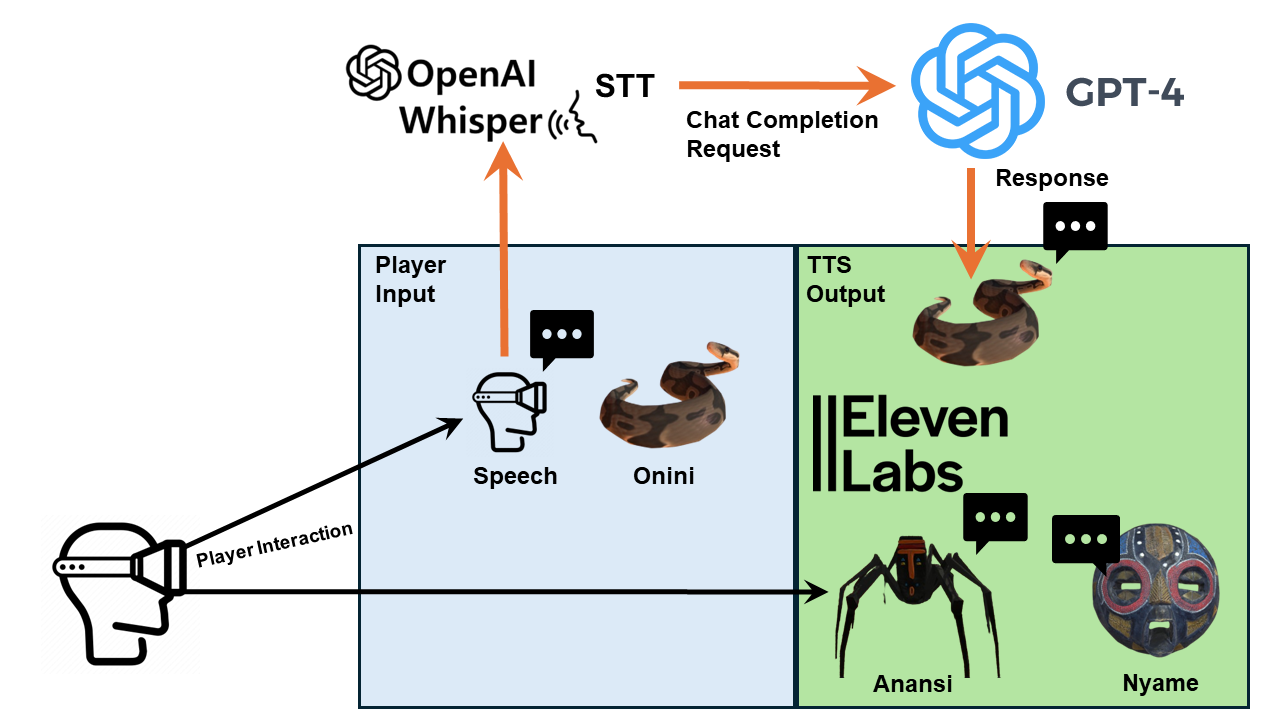}
  \caption{System Architecture of the Player NPC Interaction Model. Player speech is converted into text by OpenAI’s Whisper system, and GPT-4 generates dynamic narrative responses for the NPC Onini. The orange arrows represent the communication between Whisper, GPT-4, and ElevenLabs, where player input is processed, and responses are generated and converted back into speech. ElevenLabs handles both dynamically generated and pre-scripted text-to-speech outputs for all NPCs.}
  \Description{This diagram presents the system architecture of the 'Anansi the Spider VR' interaction model. It showcases how player interactions with non-playable characters (NPCs) are managed within the VR environment. Speech-to-text conversion is powered by OpenAI's Whisper, while GPT-4 generates dynamic narrative responses based on player input. The text-to-speech component, handled by ElevenLabs, provides real-time auditory feedback, allowing for seamless interaction between the player and the VR narrative.}
\label{fig:interaction_loop}
\end{figure}

\subsection{Measures}

The User Engagement Scale Long Form (UES-LF)~\cite{ues2018obrien}, consisting of 30 items rated from 1 (not at all) to 5 (very much), was used to measure the dependent variables: \textit{Focused Attention (FA)}, \textit{Perceived Usability (PU)}, \textit{Aesthetic Appeal (AE)}, and \textit{Reward (RW)}. The questions were randomized, and reverse coding was applied to minimize response bias. Table~\ref{tab:UES-LF_factors} in the Appendix presents the specific questions from the UES-LF used in this study. The UES-LF was chosen for its validation across various digital contexts, making it both reliable and adaptable~\cite{OBrien2010, 8668689, Bjorner2022}, and because it captures multiple dimensions of engagement. Unlike tools that focus solely on user satisfaction, the UES-LF differentiates between satisfaction and engagement, which is critical when evaluating user interactions and responses to different media or experiences.

To assess how personalization and medium differences impact cultural interests, we conducted a pre-assessment to collect data on participants' cultural backgrounds and prior exposure to storytelling technologies, including familiarity with VR, as detailed in Appendix Table~\ref{tab:Pre-Assessment}. Additionally, we used a self-designed cultural interest questionnaire, consisting of five Likert-scale questions (rated from 1, not at all interested, to 5, very interested), to measure changes in attitudes toward cultural learning and curiosity. The specific questions and scale are presented in Table~\ref{tab:CulturalInterestSurvey} in the Appendix. This approach aligns with previous research~\cite{Konstantakis2018, Souza2016, Zahidi2013}, which highlights the importance of evaluating the cultural impact on users' knowledge, emotional engagement, and connection to cultural content.

In parallel, VR interaction log data, such as play duration, walking distance, and NPC conversations, were collected using Unity. However, these data are not included in this user study, as they fall outside the scope of our research questions.

\subsection{Analysis}

\subsubsection{Quantitative Data}
Statistical analyses were conducted using Python 3.12.5\footnote{\url{https://docs.python.org/release/3.12.5/}, last accessed on 27 July 2024}. Due to violations of normality and homogeneity of variance, the Kruskal-Wallis test was applied to both UES and Cultural Interest scores. Significant findings from this test led to post-hoc Mann-Whitney U tests with Bonferroni corrections. For the Cultural Interest scores, we opted for the Holm-Bonferroni method, a less conservative approach than the traditional Bonferroni correction, to adjust for multiple comparisons.

To explore the relationships between user engagement and cultural interest further, ordinal logistic regression analysis (OLR) was conducted, guided by significant Pearson correlations between the UES and the Cultural Interest Scale. This analysis aimed to identify potential predictors of user engagement and cultural interest.

\subsubsection{Qualitative Data}
Following the methodologies suggested by Braun and Clarke~\cite{Braun_Clarke_2006} and Campbell et al.~\cite{Campbell2021ReflexiveTA}, the qualitative data from the interviews were recorded, transcribed, and subjected to thematic analysis. This process helped identify key themes such as ``Immersion,'' ``Personalization,'' ``Reflectivity,'' ``Engagement,'' and ``Cultural Relevance.'' These themes were crucial for understanding their impact on participants' overall experience and satisfaction.

To ensure the reliability of our thematic coding, we calculated Cohen's kappa for inter-rater reliability among two coders who categorized 269 items into six categories: Immersion, Personalization, Engagement, Reflection, Culture, and Technological Intervention. The kappa score is 0.72, falling within the substantial agreement range (0.61 to 0.80), as recommended by Viera and Garrett~\cite{Viera2005} and Eagan et al.~\cite{Eagan2020}, thus affirming the consistency and validity of our coding process.

\subsubsection{Researchers Positionality Statement}
In studying cultural-technical systems tied to ethnicity, such as `Anansi the Spider VR,' it is important to reflect on the author’s race, ethnicity, and the affinities they might create~\cite{Watt2015}. We are a diverse group of researchers from various fields. As the main researcher in this study, I acknowledge myself as an educated Asian researcher; my approach to studying and revitalizing folklore is informed by a background that blends academic diligence with a passion for cultural heritage. My work is driven by a commitment to understanding and preserving diverse narratives through technology. This position allows me to reinterpret folklore with a fresh perspective, aiming to make these traditional stories relevant and accessible to contemporary audiences. My values emphasize respect for cultural authenticity and the power of storytelling in bridging cultural divides.

This mixed-method approach enhanced our understanding of the complex interplay between user engagement and personalized, immersive environments. Integrating qualitative analysis allowed us to contextualize the quantitative findings more deeply, illuminating the reasons behind participants' scores on the UES and Cultural Interest Scale, thereby providing a more comprehensive understanding of user engagement dynamics.

\section{Results}

The following sections detail the findings from both the quantitative analysis and the qualitative thematic analysis to comprehensively explore how Gen-AI and VR can enrich cultural storytelling.

\subsection{User Engagement Scale (UES-LF) and Participant Interview}
Our quantitative analysis utilized the long form of the User Engagement Scale (UES)~\cite{ues2018obrien} to evaluate four dimensions of engagement: Focused Attention (\textbf{FA}), Perceived Usability (\textbf{PU}), Aesthetic Appeal (\textbf{AE}), and Reward (\textbf{RW}).

\begin{figure*}[ht]
    \centering
    \includegraphics[width=0.8\linewidth]{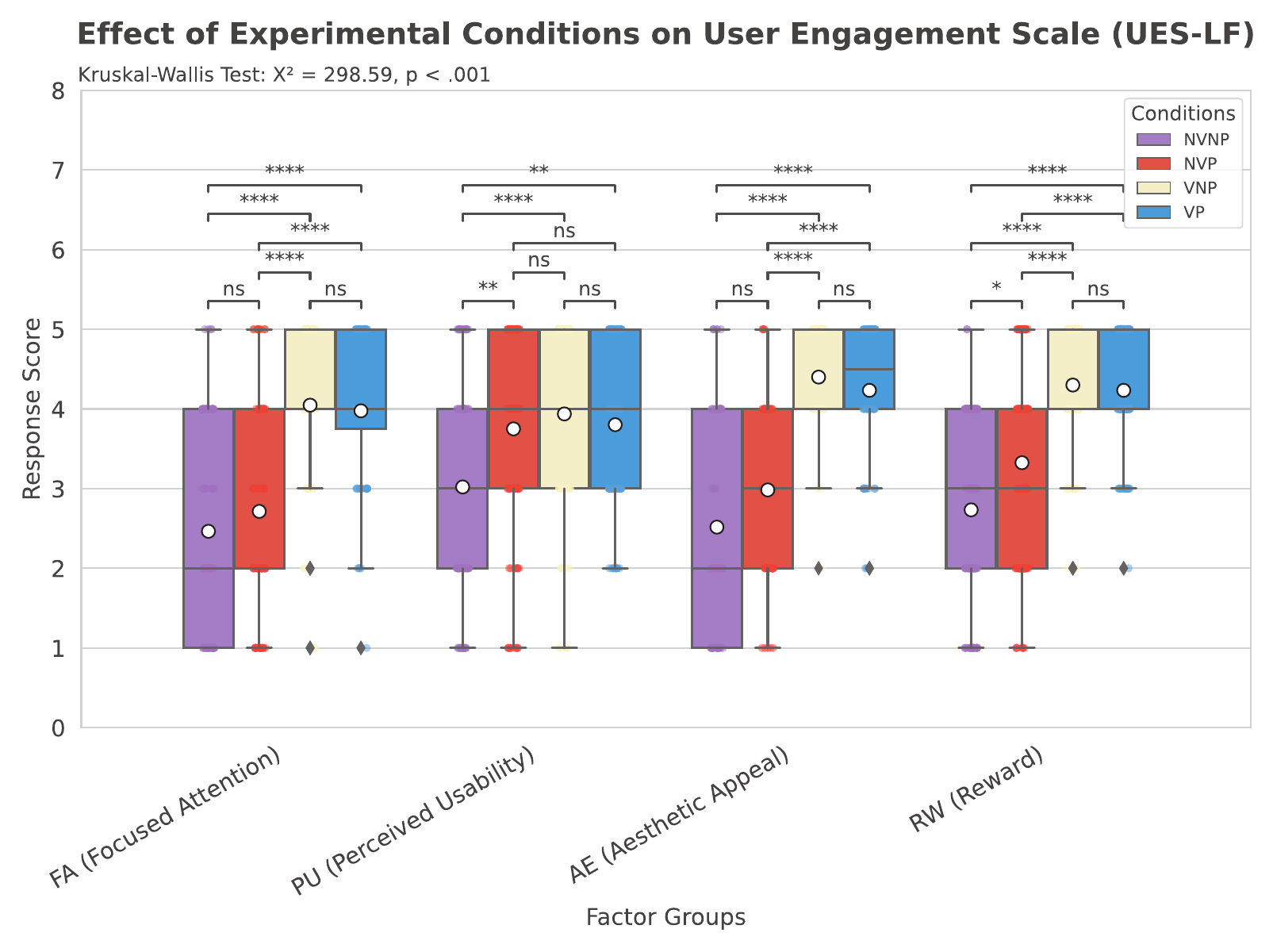}
    \caption{User Engagement Scores Comparing Different Conditions Between NVNP, NVP, VNP, and VP. *, **, ***, and **** correspond to significance levels of \( p < .05\), \( p < .01 \), \( p < .001 \), and \( p < .0001 \), respectively. Response options used a Likert scale ranging from 1 (Strongly disagree) to 5 (Strongly agree).}
    \Description{This figure compares User Engagement Scores across four different conditions: NVNP, NVP, VNP, and VP. Asterisks indicate significance levels, with * for \( p < .05\), ** for \( p < .01\), *** for \( p < .001\), and **** for \( p < .0001\). The scores are based on responses collected using a Likert scale, where participants rated their agreement with statements from 1 (Strongly disagree) to 5 (Strongly agree), reflecting their engagement in each condition.}
\label{fig:ues-box}
\end{figure*}

Figure~\ref{fig:ues-box} shows significant findings from the Kruskal-Wallis test revealed differences in UES scores across conditions (\(X^2 = 298.59\), \(p < .001\)). Post-hoc analyses using the Mann-Whitney U test with Bonferroni corrections highlighted the effectiveness of personalization, particularly in non-VR settings.

Both VR conditions (\textbf{VNP} and \textbf{VP}) generally outperformed the non-VR conditions across all dimensions, particularly in \textbf{FA}, \textbf{AE}, and \textbf{RW}, with all comparisons reaching statistical significance (\(p < .001\)). For instance, in \textbf{FA}, \textbf{VP} (\(M = 3.98\), \(SD = 0.96\)) and \textbf{VNP} (\(M = 4.05\), \(SD = 1.04\)) scored significantly higher than \textbf{NVNP} (\(M = 2.46\), \(SD = 1.26\)). In \textbf{AE}, \textbf{VP} (\(M = 4.23\), \(SD = 0.91\)) and \textbf{VNP} (\(M = 4.40\), \(SD = 0.64\)) also demonstrated greater appeal compared to \textbf{NVNP} (\(M = 2.52\), \(SD = 1.38\)).

Statistical significance was observed in the \textbf{PU} dimension, where \textbf{VP} (\(M = 3.80\), \(SD = 1.03\)) outperformed \textbf{NVNP} (\(M = 3.02\), \(SD = 1.35\), \(p = .0013\), \(U = 3094\)). Additionally, \textbf{VNP} (\(M = 3.93\), \(SD = 1.17\)) scored higher than \textbf{NVNP} (\(p < .001\), \(U = 2838\)), indicating that both VR conditions outperformed non-VR conditions. Interestingly, while \textbf{NVP} (\(M = 3.75\), \(SD = 1.27\)) was higher than \textbf{NVNP} (\(p = .0041\), \(U = 3196\)), it did not differ significantly from \textbf{VNP} (\(p = 1.00\), \(U = 4244\)).

Surprisingly, in the \textbf{RW} dimension, the \textbf{NVP} (\(M = 3.33\), \(SD = 1.24\)) vs. \textbf{NVNP} (\(M = 2.73\), \(SD = 1.18\), \(p = .01\), \(U = 5356\)) comparison showed a notable effect, suggesting that personalization within non-VR settings impacts reward perception.

To further investigate the implications of personalization when participants are exposed to both VR mediums, we utilized semi-structured interviews. These interviews allowed participants to articulate specific aspects of personalization that impacted their experience, providing rich insights that standardized scales might miss. For example, when exploring the theme of ``Engagement,'' \textbf{VP2} noted, ``VR is different because it provides more context within the story.'' This statement underscores the quantitative finding that VR significantly enhances engagement compared to non-VR settings. Additionally, personalization in VR appeared to deepen the engagement further. As \textbf{VP5} stated, ``This story will stay with me for a lifetime,'' and \textbf{VP4} commented, ``After this, I am going to Google more about the Anansi story because I want to know more.'' These reflections illustrate that personalization in VR leads to more prolonged engagement, which fosters continuous learning—a finding not observed in the non-personalized VR conditions. This contrast highlights the unique benefits of personalized experiences in immersive environments.

\subsection{Cultural Interest}
A Cultural Interest Survey was developed to assess differences in cultural engagement, changes in interest, and the applicability of the system to other traditional folklore across different countries.

\begin{figure*}[ht]
\centering
\includegraphics[width=0.9\linewidth]{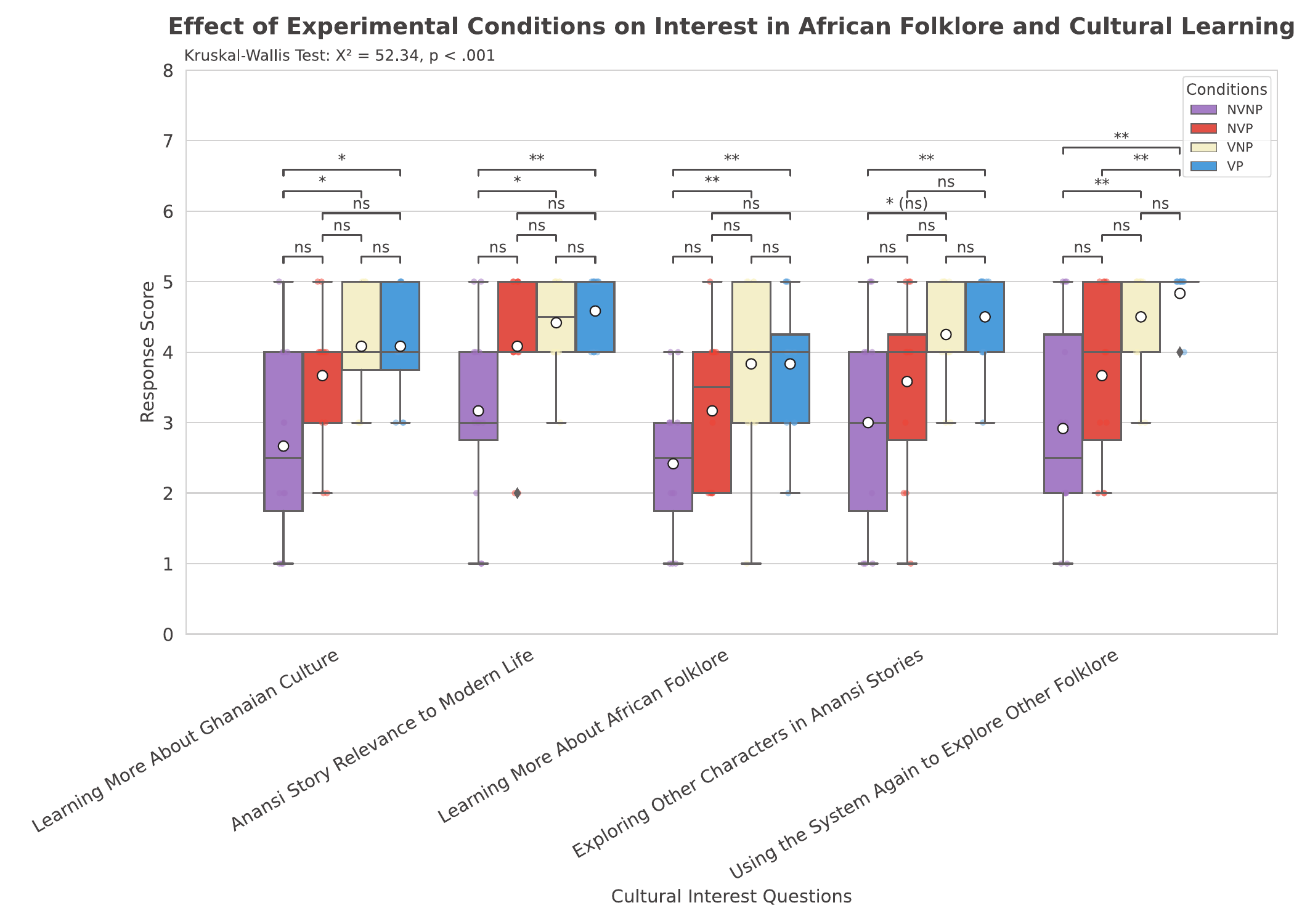}
\caption{Interest scores across different conditions (NVNP, NVP, VNP, and VP) in African Folklore and Cultural Learning. Significance levels are indicated as follows: * for \( p < .05 \), ** for \( p < .01 \), *** for \( p < .001 \), and **** for \( p < .0001 \). ``ns'' indicates non-significant differences (\( p \geq .05 \)). Response options used a Likert scale ranging from 1 (Very interested) to 5 (Not interested at all).}
\Description{This figure illustrates interest scores across four conditions: NVNP, NVP, VNP, and VP, in the domains of African Folklore and Cultural Learning. Significance levels are represented as * for \( p < .05 \), ** for \( p < .01 \), *** for \( p < .001 \), and **** for \( p < .0001 \). ``ns'' is used for non-significant differences (\( p \geq .05 \)). Participants rated their interest using a Likert scale that ranges from 1 (Very interested) to 5 (Not interested at all).}
\label{fig:culture_interest_score}
\end{figure*}

Figure~\ref{fig:culture_interest_score} present the Kruskal-Wallis test revealed significant differences in the scores of five individual cultural interest questions across the conditions (\(X^2 = 52.34\), \(p < .001\)). Subsequent post-hoc pairwise comparisons, conducted using the Mann-Whitney U test with Benjamini-Hochberg correction for multiple testing, are depicted in Figure~\ref{fig:culture_interest_score}. The results indicated consistent superior performance by VR conditions.

For \textit{``Learning More about Ghanaian Culture,''} significant differences were observed between VP (\(M = 4.08\), \(SD = 0.79\)) and NVNP (\(M = 2.67\), \(SD = 1.37\), \(p = .013\), \(U = 29.5\)), and between VNP (\(M = 4.08\), \(SD = 0.79\)) and NVNP (\(p = .013\), \(U = 29.5\)). No other significant differences were noted in this category.

For \textit{``Anansi Story Relevance to Modern Life,''} VP (\(M = 4.58\), \(SD = 0.51\)) demonstrated stronger significance compared to NVNP (\(M = 3.17\), \(SD = 1.34\), \(p = .004\), \(U = 24.5\)) than VNP (\(M = 4.42\), \(SD = 0.67\), \(p = .013\), \(U = 30.5\)).

For \textit{``Learning More about African Folklore,''} both VP (\(M = 3.83\), \(SD = 0.94\)) and VNP (\(M = 3.83\), \(SD = 1.19\)) were significantly higher than NVNP (\(M = 2.42\), \(SD = 1.08\), \(p = .005\), \(U = 24.5\) for VP and \(p = .008\), \(U = 26.5\) for VNP).

In the category of \textit{``Exploring Other Characters in Anansi Stories,''} VP (\(M = 4.50\), \(SD = 0.67\)) was significantly higher than NVNP (\(M = 3.00\), \(SD = 1.48\), \(p = .008\), \(U = 27.5\)).

For \textit{``Using the System Again to Explore Other Folklore,''} significant differences were found for VP (\(M = 4.83\), \(SD = 0.39\)) compared to NVNP (\(M = 2.92\), \(SD = 1.51\), \(p = .002\), \(U = 22.0\)), and VNP (\(M = 4.50\), \(SD = 0.80\)) compared to NVNP (\(p = .009\), \(U = 29.0\)).

No significant differences were found between VP and VNP or between NVP and NVNP, indicating that personalization within VR or non-VR mediums did not significantly impact cultural interest metrics in this study.

The qualitative insights further highlight these findings, illustrating how personalization influences more profound cultural interest. \textbf{VP6} commented, ``I think African stories are very distant from me. But this game made them so close to me. So I was more interested, even though I am not usually.'' This statement highlights the potential of using personalized narratives in VR to bridge cultural gaps and enhance engagement. Similarly, \textbf{NVP2} reflected, ``After the story, I thought to myself: Wow, there are two things about Africa I need to learn about. I will also tell my niece and nephew about this story,'' underscoring the potential of personalized content to extend cultural exchange beyond the virtual experience.

\subsection{The Relationship Between UES and Cultural Interest}

To examine the relationship between user engagement and cultural interest scores, Pearson correlation analyses were conducted. The results, presented in Appendix Figure~\ref{fig:correlation_metrics}, revealed strong positive correlations between UES scores and cultural interest (with \(r > 0.70\) for both African folklore and Ghanaian culture). Following these correlations, an Ordinal Logistic Regression (OLR) was employed to assess the predictive power of UES on cultural interest.

\begin{figure}[ht]
  \centering
  \begin{subfigure}[t]{\columnwidth}
      \centering
      \includegraphics[width=\linewidth]{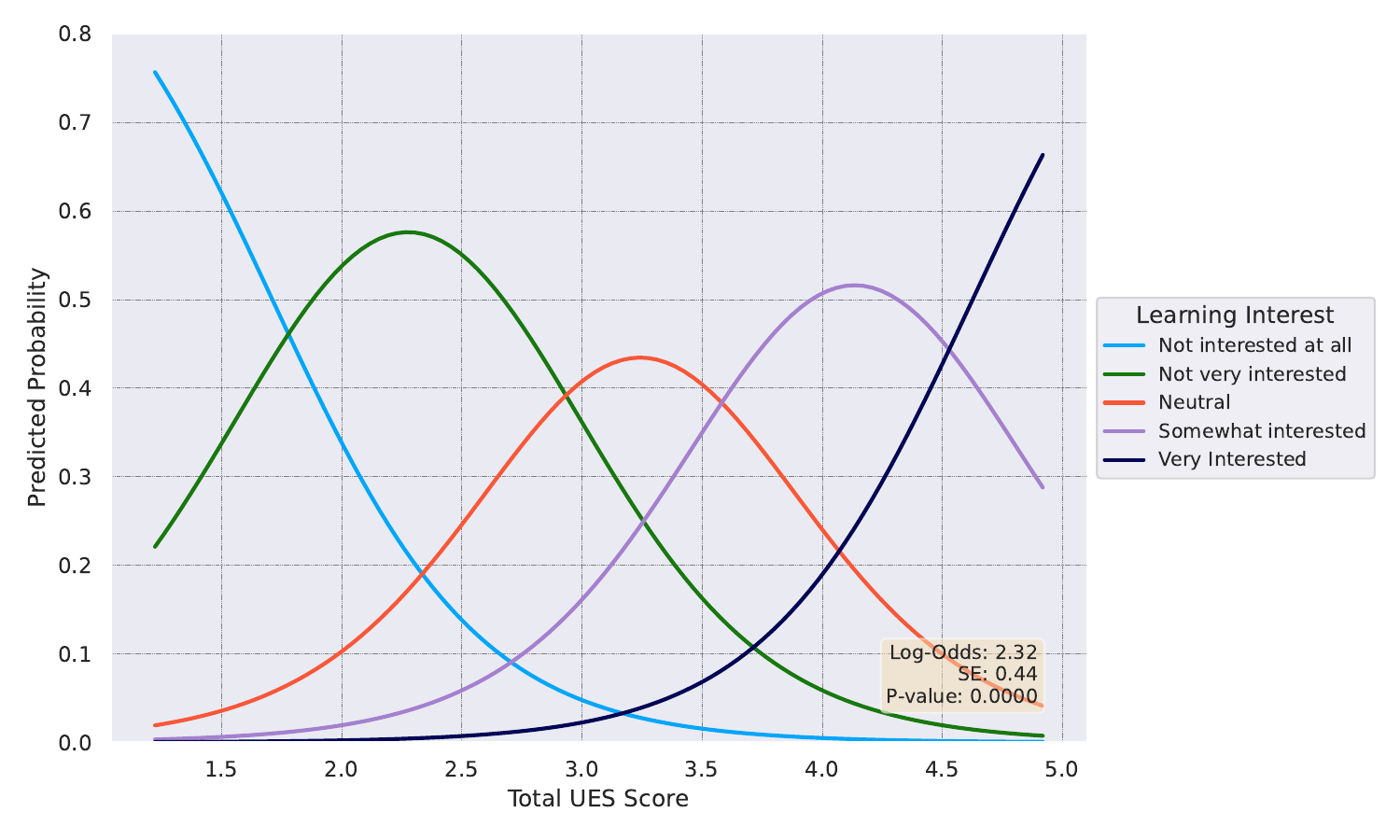}
      \caption{Predicted probabilities of learning interest in African folklore by Total UES Score based on OLR.}
      \Description{This figure shows the predicted probabilities of learning interest in African folklore based on Total UES Score using Ordinal Logistic Regression (OLR). The plot includes five categories: ``Not interested at all,'' ``Not very interested,'' ``Neutral,'' ``Somewhat interested,'' and ``Very interested.'' The predicted probabilities for each category change as the Total UES Score increases.}
      \label{fig:reg_UES_AfricanFolklore}
  \end{subfigure}
  \hfill
  \begin{subfigure}[t]{\columnwidth}
      \centering
      \includegraphics[width=\linewidth]{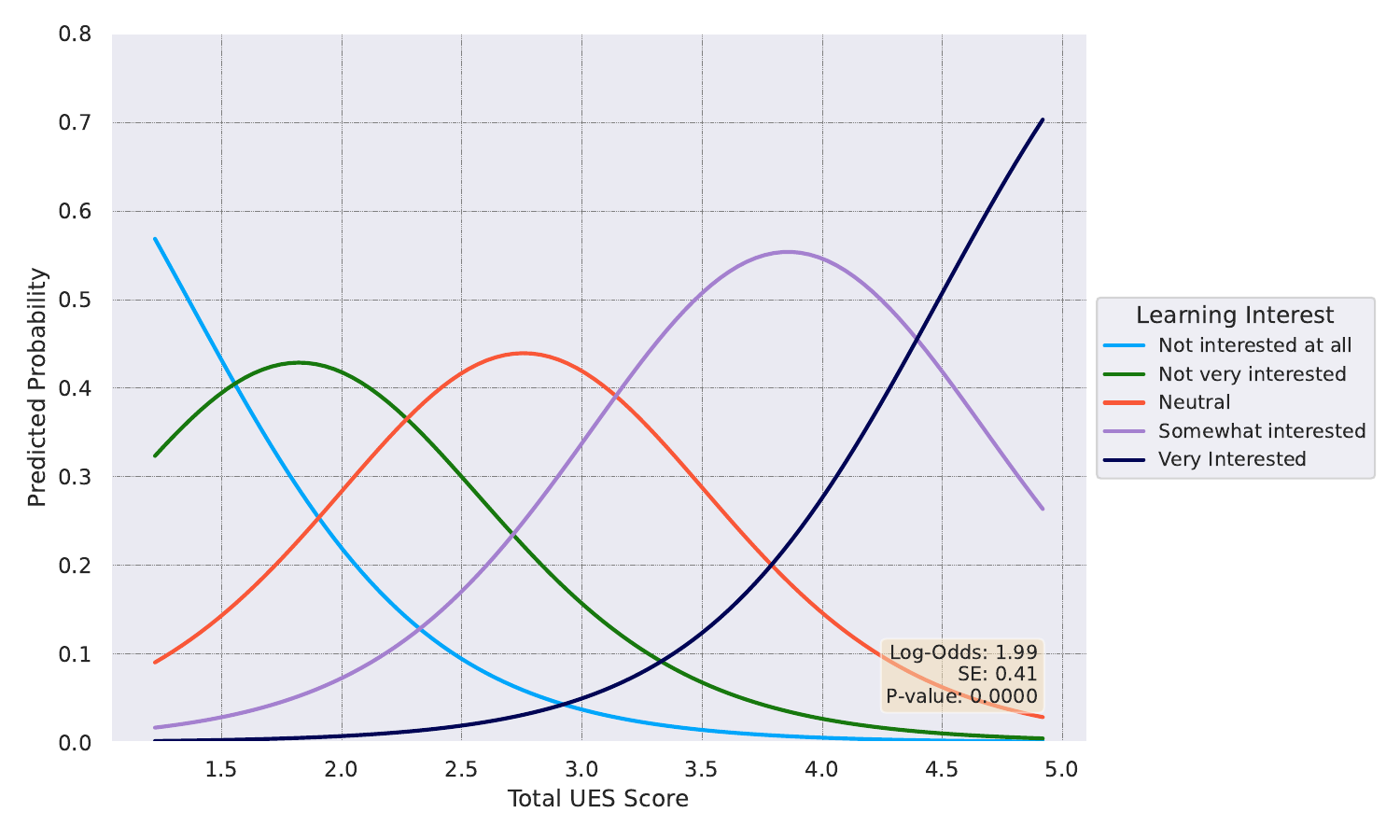}
      \caption{Predicted probabilities of learning interest in Ghanaian culture by Total UES Score based on OLR.}
      \Description{This figure presents the predicted probabilities of learning interest in Ghanaian culture based on Total UES Score using Ordinal Logistic Regression (OLR). The plot includes five learning interest categories.}
      \label{fig:reg_UES_Ghanaian}
  \end{subfigure}
  \caption{Predicted probabilities of learning interest by Total UES Score for African folklore and Ghanaian culture based on OLR.}
  \label{fig:learning_interest_OLR}
\end{figure}

As shown in Figure~\ref{fig:reg_UES_AfricanFolklore}, the predictor variable UES was found to significantly predict \textbf{learning interest in African folklore}. The ordered log-odds estimate was \(\beta = 2.32\), SE = 0.441, \(p < 0.001\). The estimated odds ratio indicated a nearly 10.19-fold increase in cultural interest (\(\text{Exp}(\beta) = 10.19\), 95\% CI [1.457, 3.186]) for every one-unit increase in UES.

Similarly, UES was also found to significantly predict \textbf{learning interest in Ghanaian culture}, as shown in Figure~\ref{fig:reg_UES_Ghanaian}. The ordered log-odds estimate was \(\beta = 1.99\), SE = 0.407, \(p < 0.001\). The estimated odds ratio indicated a 7.32-fold increase in cultural interest (\(\text{Exp}(\beta) = 7.32\), 95\% CI [1.19, 2.79]) for every one-unit increase in UES.

These findings suggest that higher UES scores are strong predictors of increased cultural interest, particularly in African folklore and Ghanaian culture. This emphasizes the role of engaging and immersive experiences in fostering cultural exploration.

\subsection{Qualitative Findings}

\begin{figure}[ht]
\centering
\includegraphics[width=\linewidth]{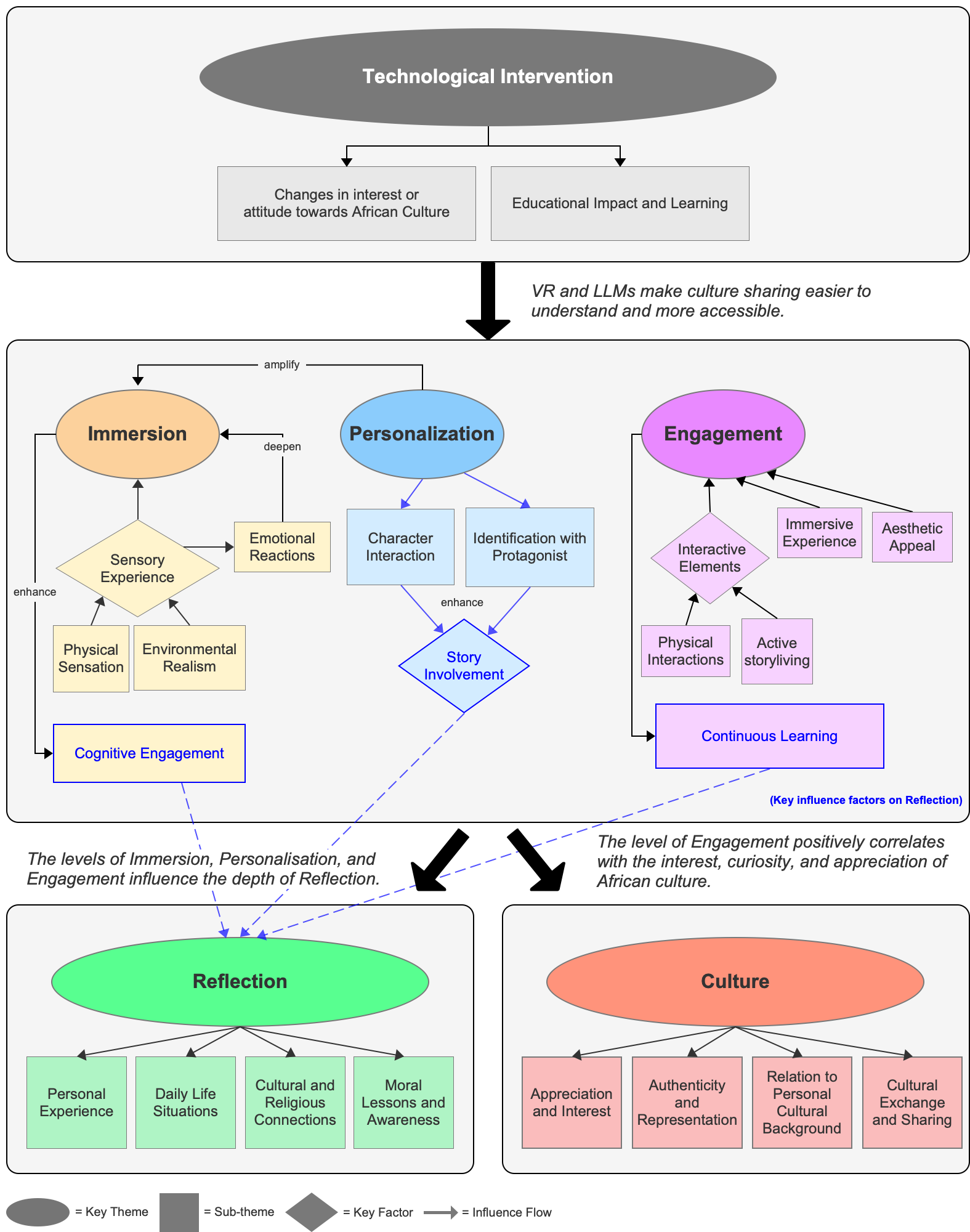}
\caption{The Role of VR and Gen-AI in Enhancing Understanding and Appreciation of African Culture: This thematic map illustrates how VR and Gen-AI influence immersion, personalization, and engagement, which in turn deepen reflection and foster a greater appreciation, authenticity, and cultural exchange related to African culture.}
\Description{This thematic map demonstrates the role of Virtual Reality (VR) and Generative AI (Gen-AI) in enhancing the understanding and appreciation of African culture. It shows how VR and Gen-AI contribute to increasing immersion, personalization, and engagement for users, which subsequently deepen reflection, promote authenticity, and facilitate cultural exchange. The map outlines the connections between these elements and their combined impact on fostering a greater appreciation for African culture.}
\label{fig:thematic}
\end{figure}

Based on our thematic analysis, we identified key themes: \textbf{Immersion}, \textbf{Personalization}, \textbf{Engagement}, \textbf{Reflection}, and \textbf{Cultural Relevance} that shaped participants' experiences in `Anansi the Spider VR.' These themes, illustrated in Figure~\ref{fig:thematic}, collectively enhanced cultural engagement and personal reflection. However, during the interviews, participants in non-VR conditions (NVP, NVNP) showed less willingness to share in-depth reflections. Their feedback often remained at surface-level observations, rarely delving into deeper cultural or personal insights. In contrast, participants in the VR conditions were more inclined to share comprehensive and meaningful reflections. The details of key themes and sub-themes identified from the qualitative analysis are reported in Appendix Table~\ref{fig:themes_immersion},~\ref{fig:themes_personalization},~\ref{fig:themes_engagement},~\ref{fig:themes_reflection},~\ref{fig:themes_culture}, and~\ref{fig:themes_intervention}.

\subsubsection{Immersion}
\textit{Enhanced Sensory and Emotional Engagement.}
Participants in the \textit{Anansi the Spider VR} experience reported a heightened sense of immersion, significantly enhancing their engagement with the narrative. Characterized by detailed \textbf{environmental realism}, this immersion made participants feel physically present within the story. For example, one participant shared, ``The jungle environment in VR was so vivid, I felt like I was about to fall off.'' highlighting the depth of immersion.

Moreover, participants experienced realistic \textbf{physical sensations}, such as a ``tickling feeling when climbing high or the tension of bending to avoid obstacles,'' which further deepened their immersive experience. The VR setup also elicited strong \textbf{emotional responses}, ranging from fear and anxiety due to lifelike dangers, to awe from the detailed visuals. As another participant noted, ``The visuals were so real and scary at times, I genuinely felt in danger, as if I were about to fall,'' underscoring the tangible and memorable nature of the experience.

The immersive experience that VR offered prompted participants to reflect on their physical and emotional responses, offering a powerful means to engage with and internalize the narrative, thus fostering a richer understanding and appreciation of the cultural content.

\subsubsection{Personalization}
\textit{Enhancing Narrative Engagement Through Personalized Interactions.}
The ability to interact within the VR environment in a personalized manner was a key factor in enhancing narrative engagement. Personalized elements such as the inclusion of participants' names significantly captured attention and fostered a sense of agency and active involvement in the story. Participants felt a stronger connection to the narrative due to the personalized nature of the experience. \textbf{VP4} remarked, ``Hearing my name during the experience made it feel as though it was designed just for me,'' which deepened engagement and made the experience feel uniquely tailored.

Particularly notable was the reaction of \textbf{VP5} from Uzbekistan, who appreciated the accurate pronunciation of their name, adding a layer of personal significance and cultural recognition: ``It pronounced my name correctly, which made the story really speak to me personally, using my real name. It gave a personalized touch that made me feel like a main character.''

\begin{quote}
``Because I was part of the story, I was more interested in what my character was doing—whether I succeeded or failed, whether I made good or bad choices. It felt more relevant and engaging.'' (NVP1)
\end{quote}

This sentiment was echoed by other participants, who felt more integral to the story's progression, transitioning from passive listeners to active participants. These responses illustrate how personalization fosters a deeper, more meaningful connection to the narrative, which is particularly impactful for individuals from diverse cultural backgrounds. Despite the absence of significant differences in engagement scores on the UES or responses to cultural interest questions, participants consistently recognized the value of personalization during the interviews. This underscores its potential to enhance narrative experiences, highlighting the importance of personalized elements even without quantifiable changes in broader engagement metrics.

\subsubsection{Engagement}
\textit{Interactivity, Active Agency, and Aesthetic Appeal are Key to Enhanced Engagement.}
Interactivity, active agency, and aesthetic appeal were identified as pivotal factors enhancing participant engagement and promoting continuous learning in the \textit{Anansi the Spider VR} experience.

\begin{quote}
``Games and stories are typically set in stone once told. However, VR experiences are valuable because they showcase how different choices can lead to varied outcomes, illustrating the real-life impact of our decisions.'' (VP10)
\end{quote}

This emphasis on interactivity was highlighted by \textbf{VP4}, who felt deeply engaged, as if truly part of the experience: ``The interaction made me feel more engaged, I was so curious I wished I could interact with everything, though it wasn’t possible... I wanted to explore every part of the world.''

Moreover, the aesthetic appeal significantly contributed to engagement. \textbf{VP5} commented on the sensory impact, ``The music and sounds were captivating; the sound of the waterfall felt incredibly real, enhancing my enjoyment.'' \textbf{VP1} supported this sentiment, noting, ``The visuals were engaging and appealing. From the beginning, it felt like you were really there. The trees and mountains were somewhat intimidating, adding to the interactivity.''

These narratives from participants illuminate the key elements that foster a robust sense of engagement when exposed to different mediums. Engagement, as described by the participants, not only keeps them involved but also facilitates continuous learning. As noted by \textbf{VP4}, ``Engaging with VR required less effort compared to traditional oral storytelling or watching a movie, and my interest grew much more easily.'' This high level of engagement encourages participants to stay immersed and more actively follow the narrative.

\subsubsection{Reflection}
\begin{figure}[ht] 
\centering \includegraphics[width=\linewidth]{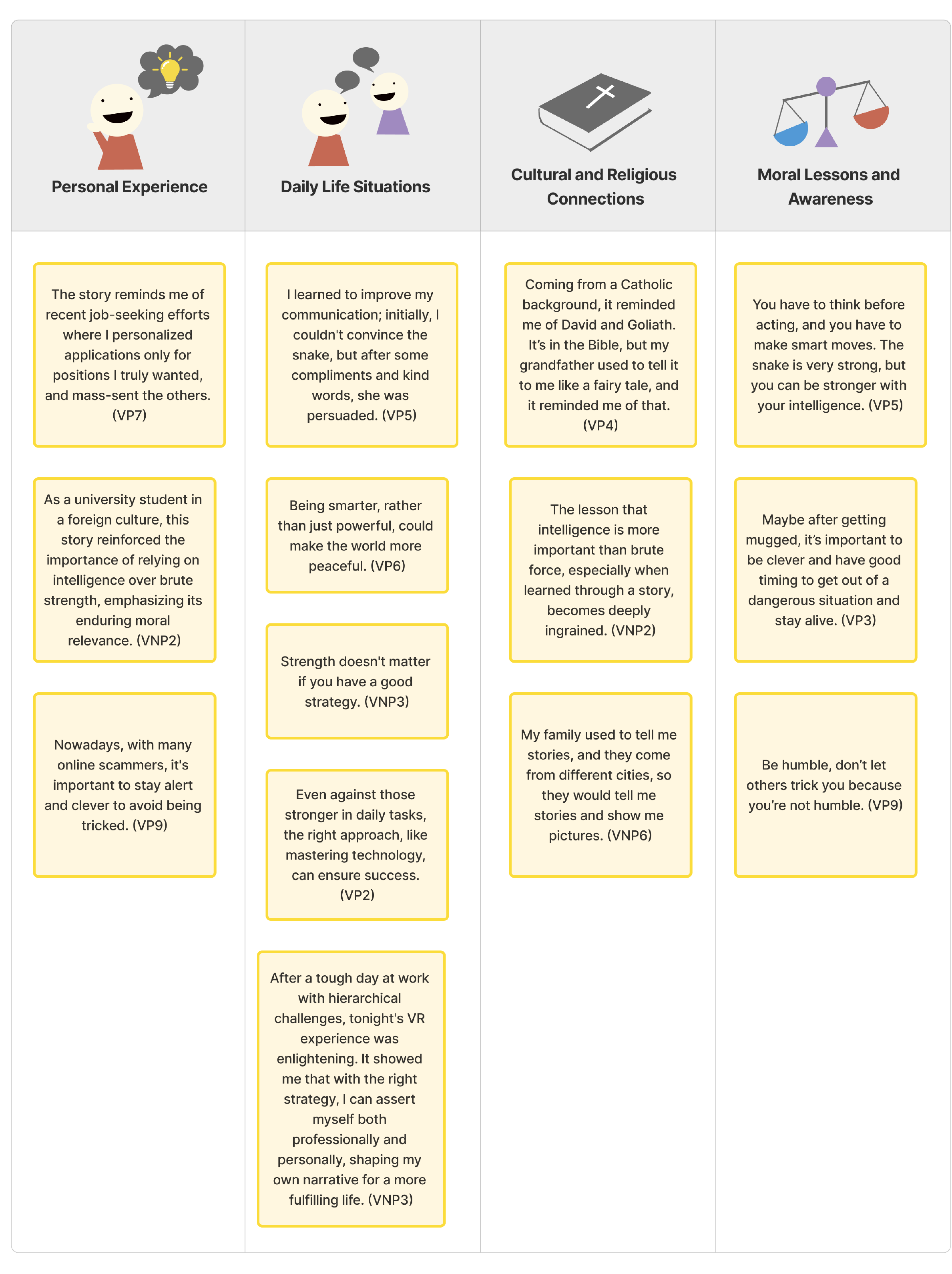}
\caption{Table of participant reflections categorized by themes: Personal Experience, Daily Life Situations, Cultural and Religious Connections, and Moral Lessons and Awareness. Each category is accompanied by relevant quotes from participants, illustrating the impact of VR and personalization on their perception and engagement with the narrative.}
\Description{This table presents participant reflections categorized into four themes: Personal Experience, Daily Life Situations, Cultural and Religious Connections, and Moral Lessons and Awareness. Each theme is supplemented with quotes from participants that highlight how VR and personalization influenced their perception and engagement with the narrative. The reflections provide insights into how the immersive experience shaped their understanding and emotional connection to the content.}
\label{fig:reflections}
\end{figure}

\textit{Involvement as the Protagonist.}
VR enabled participants to deeply reflect by immersing them in personalized roles that contrasted with their everyday lives. This immersion enhanced their sense of agency and gave them a more active role in the narrative. As participants reflected on their experiences, they revealed a dynamic process of involvement, each highlighting personal connections to the story. These reflections are categorized into four key areas, as shown in Figure~\ref{fig:reflections}. Below, we summarize the key insights from their reflection process throughout the story.

\begin{enumerate}
    \item \textbf{The Story - A Call to Adventure:} Participants frequently related their VR experiences to personal and cultural narratives, strengthening their connection to Anansi's world and its themes. \textbf{VP1} remarked, ``I was part of the story, moving from one step to the next, which deeply involved me in the unfolding events.'' Similarly, \textbf{NVP2} noted, ``It was like the Sky God assigned tasks directly to me,'' emphasizing the personalized nature of the experience.

    \item \textbf{Character Embodiment - Stepping Inside the Story:} The immersive nature of VR made participants feel as though they were embodying the characters and actively influencing the storyline. As \textbf{VP6} described, ``Being the main hero, like helping Anansi, transformed my experience from merely observing to actively influencing the narrative.''

    \item \textbf{Personal Reflection - A Hero of the Day:} Participants reflected on how their roles in the story mirrored real-life challenges. \textbf{VP7} explained how they applied narrative strategies to job applications: ``I strategized my applications like tasks in the game, focusing only on what truly mattered.'' This illustrates how VR storytelling can influence real-world decision-making. \textbf{VNP3} shared, ``After a tough day at work, tonight's VR experience showed me that with the right strategy, I can assert myself both professionally and personally, shaping my own narrative for a more fulfilling life.''

\end{enumerate}

These insights demonstrate how VR and personalized storytelling not only deepen engagement but also foster meaningful reflection, encouraging participants to reconsider their roles and potential both within and beyond the virtual world.

\subsubsection{Cultural Relevance}
\textit{Immersion, Personalization, and Engagement as Means to Revive Lost Tales.}
The combination of immersion, personalization, and engagement was crucial in enhancing the cultural relevance of the narrative. Participants appreciated the authentic representation of cultural elements, which made distant stories feel closer and sparked greater interest. \textbf{VP6} observed, ``I think African stories are usually very far away from me. But this game made them feel so close. I was more interested, even though I don't usually engage with them.''

This highlights how VR and personalization not only increased narrative engagement but also deepened participants’ appreciation for diverse cultural stories, encouraging connections to their own cultural backgrounds and promoting cultural exchange. 

\begin{quote}
``Reflecting on the story, I realized there's so much about Africa I still need to learn. I’m excited to share this narrative with my niece and nephew, hoping it sparks their interest too.'' (NVP2)
\end{quote}

These reflections suggest that cultural immersion, coupled with personalization, makes cultural narratives more reachable and fosters ongoing dialogue and exploration of diverse traditions. This experience encourages curiosity and respect for cultural diversity among participants.

\section{Discussion}
In this work, we explored the impact of Gen-AI personalization and VR immersion on oral storytelling, summarizing insights from a mixed-method approach through a user study \((n = 48)\) that assesses how these technologies influence user engagement and interest in cultural learning.

Addressing \textbf{RQ1}, findings from both the UES and qualitative interviews indicate that Gen-AI-driven personalization can enhance user engagement and foster more profound reflection on the narratives. In non-VR settings, quantitative data showed significant improvements in Perceived Usability (PU) and Reward (RW) on the UES with personalization, notably higher in the personalized group compared to non-personalized. On the contrary, in VR settings, the effect of personalization was subtler, likely overshadowed by the novelty effects of VR technology, a common phenomenon identified by previous studies~\cite{miguel2024evaluation, 9388893}. Addressing \textbf{RQ2}, the UES and Cultural Interest Questionnaire data consistently revealed that VR significantly boosts user engagement and enhances interest in cultural learning and exploration, aligning with existing literature on immersive storytelling's effectiveness~\cite{Tong2024, YuZhiyuan2023, Pillai2019}.

In the subsequent sections, we discuss the implications of our findings on the synergy between personalization and VR immersion for cultural preservation, bridging cultural distances, and enhancing personal reflection and moral awareness. We also explore how these traditional tales can be revitalized in the digital era by presenting the narrations (the artifacts) generated by our participants throughout the experiment, documented in Appendix Figure~\ref{fig:storyGeneration}. We propose design recommendations for balancing authenticity with innovation and conclude by discussing the limitations and future research directions in this dynamic field.

\subsection{Synergy of Gen-AI Personalization and VR Immersion in Storytelling}
Through this research, one of the main findings is that immersion and personalization closely complement each other. We address \textbf{RQ3} in three directions: cultural preservation, bridging cultural distances through understanding minority cultures and enhancing personal perspectives on self-reflection and moral awareness.

\subsubsection{Cultural Preservation: Enhancing Engagement for Cultural Transmission}

Our study demonstrates that incorporating personalization in VR storytelling enhances cultural engagement and supports the transmission of folklore. This finding aligns with previous research that highlights the importance of correctly pronouncing an individual’s name in education~\cite{Jane2024}, human-robot interaction~\cite{Kennedy2024}, and intercultural communication~\cite{Dali2022}. Similarly, our user study shows that personalized VR experiences, such as the Anansi story, make cultural narratives more relatable.

Participants felt a stronger connection when their names were integrated into the story and when they engaged with personalized, Gen-AI-powered interactivity. This dynamic interaction allowed them to make decisions and influence the narrative, further motivating them to explore and overcome challenges in the VR environment. Enhanced personalization not only deepened their understanding of the cultural elements but also bridged the gap between their own backgrounds and African folklore, fostering curiosity and engagement with the cultural context.

Our qualitative findings reveal that VR experiences can enhance engagement and facilitate cultural transmission, not only with foreign cultures but also within a participant's own cultural heritage. As \textbf{VP10} from Ghana noted,``We did not live much with our extended family, so the oral culture might have faded at that time.'' This experience offers a solution by creating virtual spaces that reconnect past traditions with the present, rekindling interest in cultural heritage among younger generations and those who may feel disconnected from their roots.

Our findings confirm that personalized VR experiences promote cultural transmission. As discussed in Section 4, participants reported feeling more engaged and expressed their intention to share what they had learned with their families. Figure~\ref{fig:thematic} illustrates how personalization increases story involvement, while immersive experiences enhance cognitive engagement, contributing to the preservation and transmission of cultural heritage.

\subsubsection{Bridging Cultural Distances: Fostering Cultural Inclusion}

Findings from our mixed-method approach indicate that personalization and VR effectively transform users from passive listeners into active participants in cultural storytelling. Previous studies have also highlighted how engaging participants in the storytelling process can promote cultural inclusion~\cite{Milano2023, di2023serious}.

Choo et al. note that traditional storytelling often creates a separation between the audience and the cultural context, emphasizing how technological interventions can enhance the storytelling experience~\cite{Choo2020}. In contrast to traditional storytelling, our study explores how Gen-AI-driven personalization fosters a more inclusive storytelling experience. This approach aligns with Hand and Varan’s research, which emphasizes that `Interactivity' within a narrative can fosters users' empathy~\cite{Hand2009}. 

Our findings demonstrate that personalization and immersion make stories from different cultures more relatable. This is reflected in both quantitative and qualitative results, with participants in the VP condition scoring higher in categories such as ``Anansi Story Relevance to Modern Life'', ``Interest in Exploring Other Characters in Anansi Stories'', and ``Willingness to Use the System Again to Explore Other Folklore''. Qualitatively, \textbf{VP6} noted, ``This game made African stories, which once felt distant, feel much closer and more engaging.'' The synergy between VR’s immersive qualities and AI-driven personalization enables a deeper connection to unfamiliar cultural elements, effectively bridging cultural distances.

\subsubsection{Enhanced Reflection Through Immersive Storytelling}

The immersive nature of VR deepens participants' connection to the narrative and fosters transformative experiences that enhance emotional engagement. Previous studies have demonstrated VR's potential to encourage reflection on topics such as death, loneliness~\cite{bahng2020reflexive}, and self-awareness~\cite{Wagener2023}. In this work, we situate VR immersiveness within the context of reflecting on oral traditions, specifically exploring how interactivity contributes to the reflective process.

By placing users in a vivid, interactive narrative environment, they transition from passive listeners to active contributors. This aligns with Liu et al.'s concept of `Embodied Storytelling,' which emphasizes VR as an effective educational tool. Through interaction and participation, VR fosters meaningful reflections, guiding participants to engage with cultural and moral lessons embedded within the narrative~\cite{Liu2022}.

In addition to immersion, we allowed users to interact with a character (Onini), introducing personalized narration that placed participants in ethical decision-making scenarios—such as baiting Onini. This interaction deepened participants' personal connection to the content, encouraging them to reflect not only on the story but also on their values and decision-making processes. This aligns with previous research on `Affective Experience' in storytelling, which shows how immersive and interactive elements work together to create emotional weight in the experience, driving deeper reflection~\cite{Maud2021, Ostrin2018, maxwell2014}.

While quantitative results were unable to fully capture the abstract feelings of reflection, we adopted semi-structured interviews to explore this process further. Through thematic analysis, we categorized participants' reflections into four types, as illustrated in Figure~\ref{fig:reflections}. Participants in the VR conditions were not only more willing to share their reflections, but the depth of their insights also expanded to include cultural, religious, and moral awareness.

Additionally, the choice of using Anansi the Spider stories proved effective, as Anansi, though small compared to the giant animals in the jungle, uses versatility and intellect to overcome brute strength. This theme resonated with participants, many of whom reflected on their own experiences of feeling like the ``hero of the day.'' They identified with the idea that success and victory do not always require physical power but can come from cleverness, problem-solving, and adaptability.

\subsection{Revitalizing Folklore in the Digital Era}

Integrating Gen-AI as a character in our VR experience demonstrates how traditional narratives can be both preserved and expanded, bridging historical heritage with modern expression. Participants actively engaged with the story, employing different strategies to persuade the snake while some succeeded and others faced rejection. This interaction, detailed in Appendix Figure~\ref{fig:storyGeneration}, illustrates how AI-driven personalization invites deeper user involvement and co-creation.

This approach aligns with Sillett's concept of `Modern Orality,' which emphasizes the role of digital platforms in preserving folklore while allowing for creative adaptations~\cite{sillett2020folklore}. By allowing audiences to shape and participate in the narrative, we enable these tales to evolve while maintaining their cultural relevance, in line with Toelken's perspective that folklore must balance both static and changing elements to remain meaningful in modern contexts~\cite{toelken1996dynamics}.

While the development of Gen-AI is still being challenged~\cite{yao2024llmlieshallucinationsbugs, Fan2024}, it is clear that it offers opportunities for fostering creativity and personal connections within storytelling. As discussed in~\cite{Fan2024, FuKexue2024}, the co-creation of digital narratives promotes deeper reflection, emotional engagement, and sustained cultural discourse. Our findings suggest that integrating AI within VR not only revitalizes folklore but also offers new avenues for creative expression and community-driven engagement with cultural heritage.

\subsection{Design Recommendations for Cultural Preservation and Transmission}

Combining our findings and prior literature, we propose two design recommendations for creating personalized VR interventions that balance cultural authenticity while fostering cultural reflections and transmission.

\subsubsection{Recommendation 1: Maintain Story Authenticity in Interactive Narrative Structures}

When designing immersive narratives based on oral traditions, it is essential to balance innovation with the preservation of the story's core structure. While modern audiences expect interactive elements in VR experiences, these should not distort the original narrative’s cultural meaning. Involving cultural insiders during the development process ensures the cultural essence is maintained even as the story offers interactive choices. 

For instance, in `Anansi the Spider VR,’ while participants were addressed personally, the core narrative adhered closely to the original tale. Participants appreciated this balance, as seen with \textbf{VP4}, who expressed a preference for choices that maintained the story’s ultimate outcome: ``I like when stories are like, you have A and B, and regardless of A or B, the outcome is the same.'' This sentiment was echoed by \textbf{VP9}, who emphasized: ``It's normal to change stories, but maybe it's also nice to not change too much so the original story can be passed on.''

Our approach used a ``yo-yo'' narrative model~\cite{Hand2009}, where participants could explore alternate branches while always returning to the main plot. This technique preserves narrative cohesion while allowing meaningful interaction. Future immersive storytelling should similarly preserve the structural integrity of traditional tales while offering opportunities for user-driven interaction.

\subsubsection{Recommendation 2: Align VR Experiences with Cultural Origins and Context}

In addition to maintaining narrative structure, immersive VR experiences should accurately reflect the cultural context from which the oral tradition originates. Oral traditions are deeply connected to the environments in which they were told, such as the communal spaces where stories were shared around a campfire. \textbf{VP8} highlighted this need for authenticity: ``More visuals to show this was how our culture started, and where it started from, and then do the story.''

This points to a need for VR experiences to ground users in the cultural and historical origins of the narrative. Jones’ research~\cite{Jones2017} warns that VR’s visual capabilities can bias the storytelling by framing the narrative in ways that might overshadow its core message. Thus, care must be taken in how the story is introduced and framed within the VR experience, starting from culturally significant settings that resonate with the tradition.

Creators should collaborate with cultural insiders to ensure the VR environment—such as the setting of a story around a campfire—is portrayed authentically. This will help users feel more connected to the cultural roots of the narrative. Moreover, while modern technology allows for interactivity, creators should prioritize preserving key cultural moments and settings, ensuring that the immersive experience reflects the tradition’s original context.

\section{Limitations and Future Work}

\textit{Standardized engagement scales face challenges in interpreting personal experiences.} 
A key challenge encountered in this study was the limited sensitivity of quantitative results for RQ1. High UES scores for VR conditions suggest a ceiling effect, which may have masked subtle differences between personalized and non-personalized experiences. These nuanced differences were better captured in semi-structured interviews, where participants offered more detailed reflections. This discrepancy between qualitative insights and quantitative data highlights the limitations of standardized scales like UES in capturing emotional and cognitive depth, underscoring the importance of mixed-method approaches.

\textit{Incorporating think-aloud methods could provide real-time insights into user experiences.} To gain deeper insights into user experiences, future research could use the think-aloud method during VR sessions. This approach provides real-time access to participants' thoughts and emotions, offering richer qualitative data beyond traditional metrics. Although think-aloud methods are rarely used in VR studies, Siette et al.~\cite{Siette2024} applied them to assess VR usability in older adults. While some argue that this disrupts immersion~\cite{winn2002does}, guided verbal reflection could enrich understanding, especially in complex VR scenarios. Capturing these reflections may reveal hidden effects of VR and AI on cultural learning.

\textit{Longitudinal studies are essential for assessing long-term cultural impact.} To evaluate the sustained effects of VR and Gen-AI on cultural engagement, future research can adopt longitudinal approaches, such as long-term participatory design. On-going engagement with communities could reveal participants' evolving needs and perceptions, offering valuable insights into how these technologies foster lasting cultural connections and influence real-world behavior~\cite{Konstantakis2018, Giglitto2019}. By incorporating longitudinal methods, researchers could better understand how VR and Gen-AI contribute to cultural preservation over time.

\textit{Public collaboration with cultural institutions is essential for effective preservation.} Otero-Pailos emphasizes the need for innovative approaches in preservation, urging cultural institutions to re-think traditional methods of conserving historical and cultural heritage~\cite{Otero-Pailos2016-xv}. In the future, we plan to collaborate with public institutions, such as museums, to evaluate the effectiveness of using VR and Gen-AI for more dynamic, participatory preservation methods. Rather than viewing preservation as keeping artifacts unchanged, these technologies allow cultural narratives to evolve, involving the public in storytelling. This collaboration could transform the preservation process from an expert-led narrative into one shaped by a broader, more inclusive community.

\section{Conclusion}

To investigate how VR immersion and Gen-AI-driven personalization can enhance storytelling, particularly in the context of preserving oral traditions, we developed `Anansi the Spider VR', an innovative approach to revitalizing a Ghanaian oral tradition. Through a mixed-methods design with \((n = 48)\) participants, we explored the impact of VR immersion and Gen-AI-driven personalization on engagement and cultural learning. Quantitative findings showed that both immersion and personalization significantly enhanced user engagement and interest in cultural learning.

Qualitative findings revealed that VR and Gen-AI can transform user participation in storytelling, fostering cultural transmission, bridging cultural distances, and promoting self-reflection. Participants not only connected with African folklore but also reflected on their own experiences through the lessons of the stories. These insights highlight the role of immersive technologies in preserving and revitalizing cultural narratives.

This study also identified challenges in quantifying personal experiences within VR, underscoring the limitations of standardized tools like the UES. Future research should focus on developing more nuanced measurement methods and collaborating with cultural institutions to further explore and preserve immersive storytelling experiences.

Overall, this research underscores the potential of VR and Gen-AI to maintain the relevance and impact of traditional tales in the digital age, ensuring that stories like Anansi continue to inspire and educate future generations.

\bibliographystyle{ACM-Reference-Format}
\bibliography{main}

\onecolumn
\appendix
\section{Researchers Positionality Statement}
As the main researcher in this study, I acknowledge myself as an educated Asian researcher; my approach to studying and revitalizing folklore is informed by a background that blends academic diligence with a passion for cultural heritage. My work is driven by a commitment to understanding and preserving diverse narratives through technology. This position allows me to reinterpret folklore with a fresh perspective, aiming to make these traditional stories relevant and reachable to contemporary audiences. My values emphasize respect for cultural authenticity and the power of storytelling in bridging cultural divides.

\section{Pre-Assessment Questionnaire}

\begin{table}[ht]
  \caption{Pre-Assessment Questionnaire: Items and Response Options for Evaluating Participants' Background, Familiarity, and Preferences.}
  \Description{This table presents the items and response options from the pre-assessment questionnaire used to evaluate participants' backgrounds, familiarity with certain topics, and preferences. The questionnaire covers a range of questions about cultural background, education level, familiarity with digital storytelling, virtual reality, and oral traditions, as well as interest in cultural learning. Participants responded using predefined scales or multiple-choice options depending on the question.}
  \label{tab:Pre-Assessment}
  \centering
  \begin{tabular}{p{10cm}p{7cm}} % Adjusted column widths for a single-column format
    \toprule
    \textbf{Questions} & \textbf{Response Options} \\
    \midrule
    How would you describe your cultural background? (e.g., ethnicity, cultural upbringing, media from cultures you consumed a lot growing up) & Open-ended \\
    \midrule
    How would you describe your education level? & High School, Bachelor's Degree, Master's Degree, Doctoral Degree \\
    \midrule
    How often do you engage with digital storytelling (e.g., audiobooks, podcasts, interactive stories)? & Never, Rarely, Sometimes, Often, Very Often \\
    \midrule
    How familiar are you with virtual reality technology? & Not at all familiar, Slightly, Moderately, Very, Extremely \\
    \midrule
    How familiar are you with the concept of oral traditions? & Not at all familiar, Slightly, Moderately, Very, Extremely \\
    \midrule
    How familiar are you with African folklore? & Not at all familiar, Slightly, Moderately, Very, Extremely \\
    \midrule
    How interested are you in learning about other cultures? & Not at all interested, Slightly, Moderately, Very, Extremely interested \\
    \midrule
    How comfortable are you with technology adapting content based on your personal information? & Very uncomfortable, Uncomfortable, Neutral, Comfortable, Very Comfortable \\
    \midrule
    Which methods do you prefer for learning about cultural heritage? (Select all that apply) & Reading books and articles, Museums and historical sites, Interactive experiences (workshops, video games), Watching films and documentaries, Online resources (websites, blogs), Travelling, Talking to family and friends \\
    \bottomrule
  \end{tabular}
\end{table}

\clearpage
\section{User Engagement Scale Long Form}    

\begin{table}[ht]
  \caption{Summary of Factors and Items from UES-LF Questionnaire, with Responses Rated on a Likert Scale from 1 (Strongly Disagree) to 5 (Strongly Agree).}
  \Description{This table summarizes the factors and items from the UES-LF (User Engagement Scale-Long Form) questionnaire. Participants rated their responses on a Likert scale ranging from 1 (Strongly Disagree) to 5 (Strongly Agree). The table includes various factors and corresponding items that measure different aspects of user engagement.}
  \label{tab:UES-LF_factors}
  \centering
  \begin{tabular}{p{3cm}p{2cm}p{10cm}} % Adjust column widths
    \toprule
    \textbf{Factors} & \textbf{Items} & \textbf{Questions} \\
    \midrule
    FA     & FA.1  & I lost myself in the Anansi story experience. \\
           & FA.2  & I was so involved in the Anansi story experience that I lost track of time. \\
           & FA.3  & I blocked out things around me using the Anansi story experience. \\
           & FA.4  & Using the Anansi story experience, I lost track of the world around me. \\
           & FA.5  & The time I spent using the Anansi story experience just slipped away. \\
           & FA.6  & I was absorbed in this experience. \\
           & FA.7  & During this experience I let myself go. \\
    \midrule
    PU     & PU.1  & I felt frustrated while using the Anansi story experience. \\
           & PU.2  & I found the Anansi story experience confusing to use. \\
           & PU.3  & I felt annoyed while using the Anansi story experience. \\
           & PU.4  & I felt discouraged while using the Anansi story experience. \\
           & PU.5  & Using the Anansi story experience was taxing. \\
           & PU.6  & This experience was demanding. \\
           & PU.7  & I felt in control while using the Anansi story experience. \\
           & PU.8  & I could not do some of the things I needed to do while using the Anansi story experience. \\
    \midrule
    AE     & AE.1  & The Anansi story experience was attractive. \\
           & AE.2  & The Anansi story experience was aesthetically appealing. \\
           & AE.3  & I liked the graphics and images of the Anansi story experience. \\
           & AE.4  & The Anansi story experience appealed to my visual senses. \\
           & AE.5  & The screen layout of the Anansi story experience was visually pleasing. \\
    \midrule
    RW     & RW.1  & Using the Anansi story experience was worthwhile. \\
           & RW.2  & I consider my experience a success. \\
           & RW.3  & This experience did not work out the way I had planned. \\
           & RW.4  & My experience was rewarding. \\
           & RW.5  & I would recommend the Anansi story experience to my family and friends. \\
           & RW.6  & I would continue using the Anansi story experience. \\
           & RW.7  & The content of the Anansi story experience incited my curiosity. \\
           & RW.8  & I was really drawn into this experience. \\
           & RW.9  & I felt involved in this experience. \\
           & RW.10 & This experience was fun. \\
    \bottomrule
  \end{tabular}
\end{table}

\clearpage
\section{Cultural Interest Questionnaire}  

\begin{table}[ht]
  \caption{Summary of Cultural Interest Questionnaire Items and Likert Scale.}
  \Description{Summary of Cultural Interest Questionnaire Items and Likert Scale.}
  \label{tab:CulturalInterestSurvey}
  \centering
  \begin{tabular}{p{10cm}p{5cm}} % Adjusted column widths for single-column format
    \toprule
    \textbf{Questions} & \textbf{Likert Scale (1-5)} \\
    \midrule
    How interested are you in learning more about Ghanaian culture after reading the Anansi story? & Not interested at all to Very interested \\
    \midrule
    How relevant do you think the moral lessons from the Anansi story are to modern life? & Not relevant at all to Very relevant \\
    \midrule
    How likely are you to seek out more African folklore after reading the Anansi story? & Not likely at all to Very likely \\
    \midrule
    Anansi stories feature other interesting characters, such as Tiger, Turtle, and Monkey. How interested are you in learning about these characters and their stories? & Not interested at all to Very interested \\
    \midrule
    After the user study, would you be interested in using the same system to learn about oral traditions from other cultures, such as Grimm’s Fairy Tales? & Not interested at all to Very interested \\
    \bottomrule
  \end{tabular}
\end{table}

\section{User-generated Narrations}

\begin{figure*}[ht]
\centering
\includegraphics[width=0.7\linewidth]{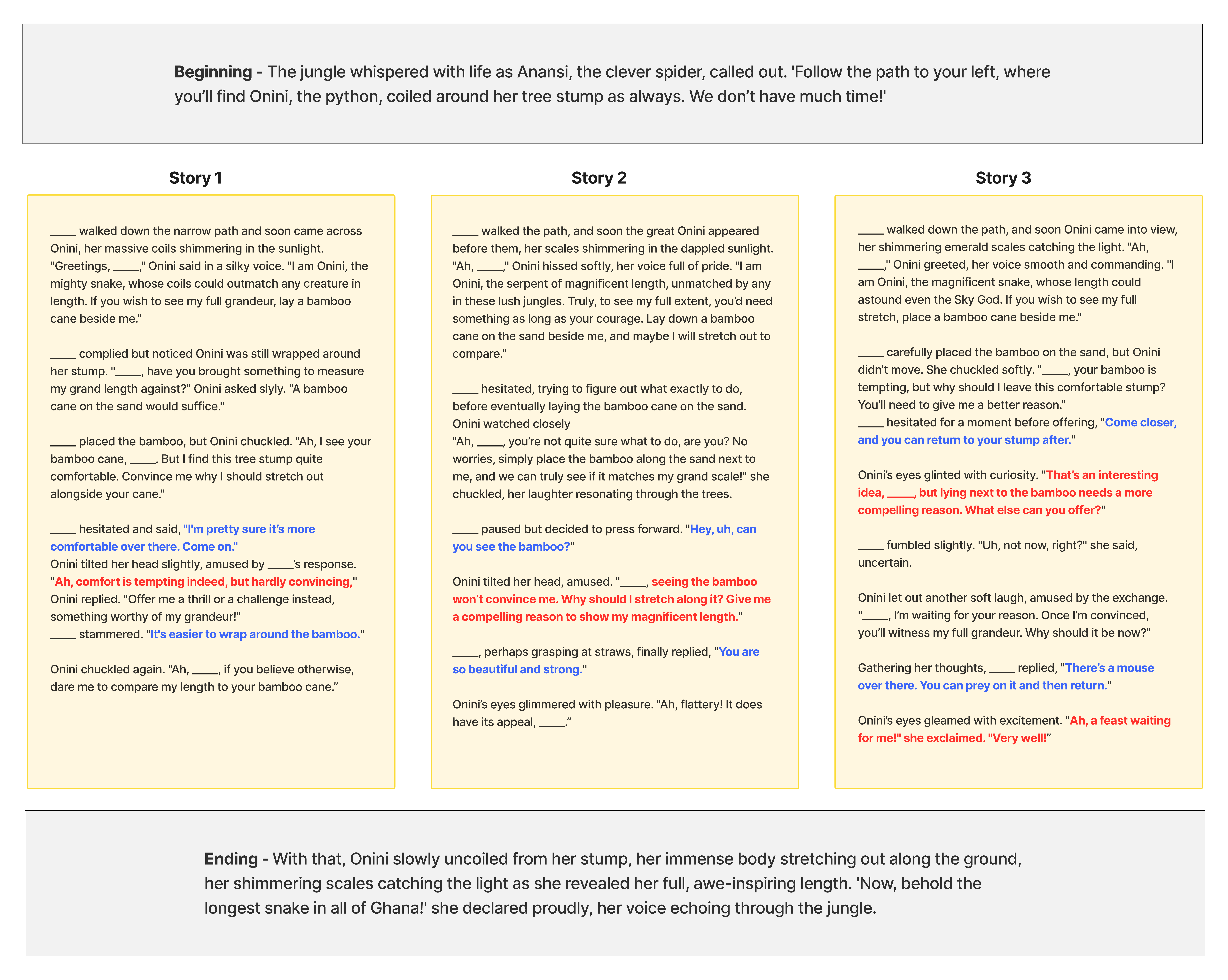}
\caption{Figure illustrates three user-generated stories in Anansi the Spider VR. Blue denotes user inputs via speech-to-text, actively shaping the tale, while red illustrates AI-generated responses via text-to-speech, demonstrating the dynamic interplay between participant choices and Gen-AI storytelling.}
\Description{This figure illustrates three user-generated stories in Anansi the Spider VR. Blue denotes user inputs via speech-to-text, actively shaping the tale, while red illustrates AI-generated responses via text-to-speech, demonstrating the dynamic interplay between participant choices and Gen-AI storytelling.}
\label{fig:storyGeneration}
\end{figure*}

\newpage

\section{Thematic Tables}

\begin{figure*}[ht]
  \centering
  \includegraphics[width=\linewidth]{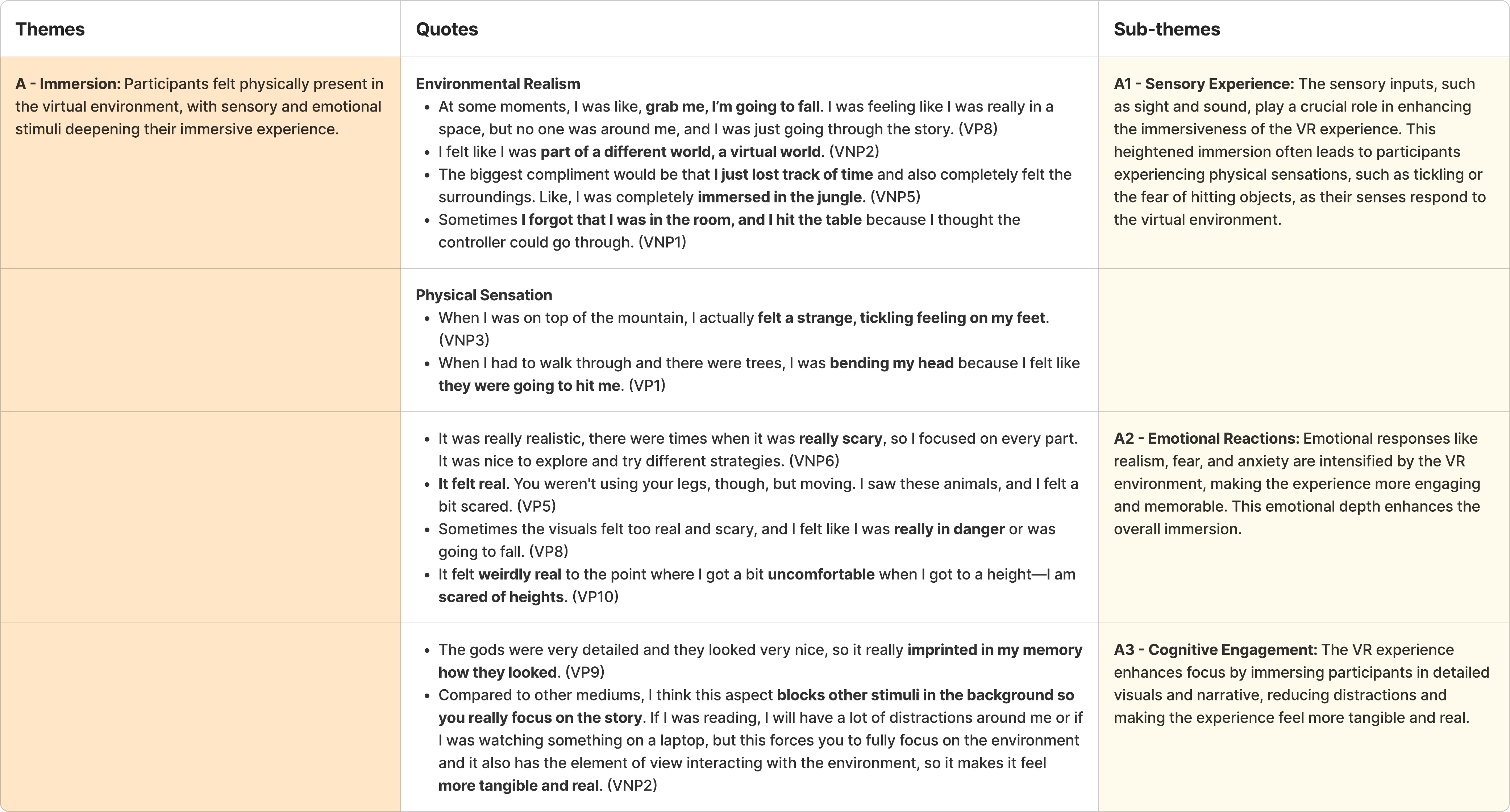}
  \caption{Thematic analysis results: Immersion}
  \label{fig:themes_immersion}
  \Description{This table presenting the theme Immersion, its sub-themes, and representative quotes from participants.}
\end{figure*}

\begin{figure*}[ht]
  \centering
  \includegraphics[width=\linewidth]{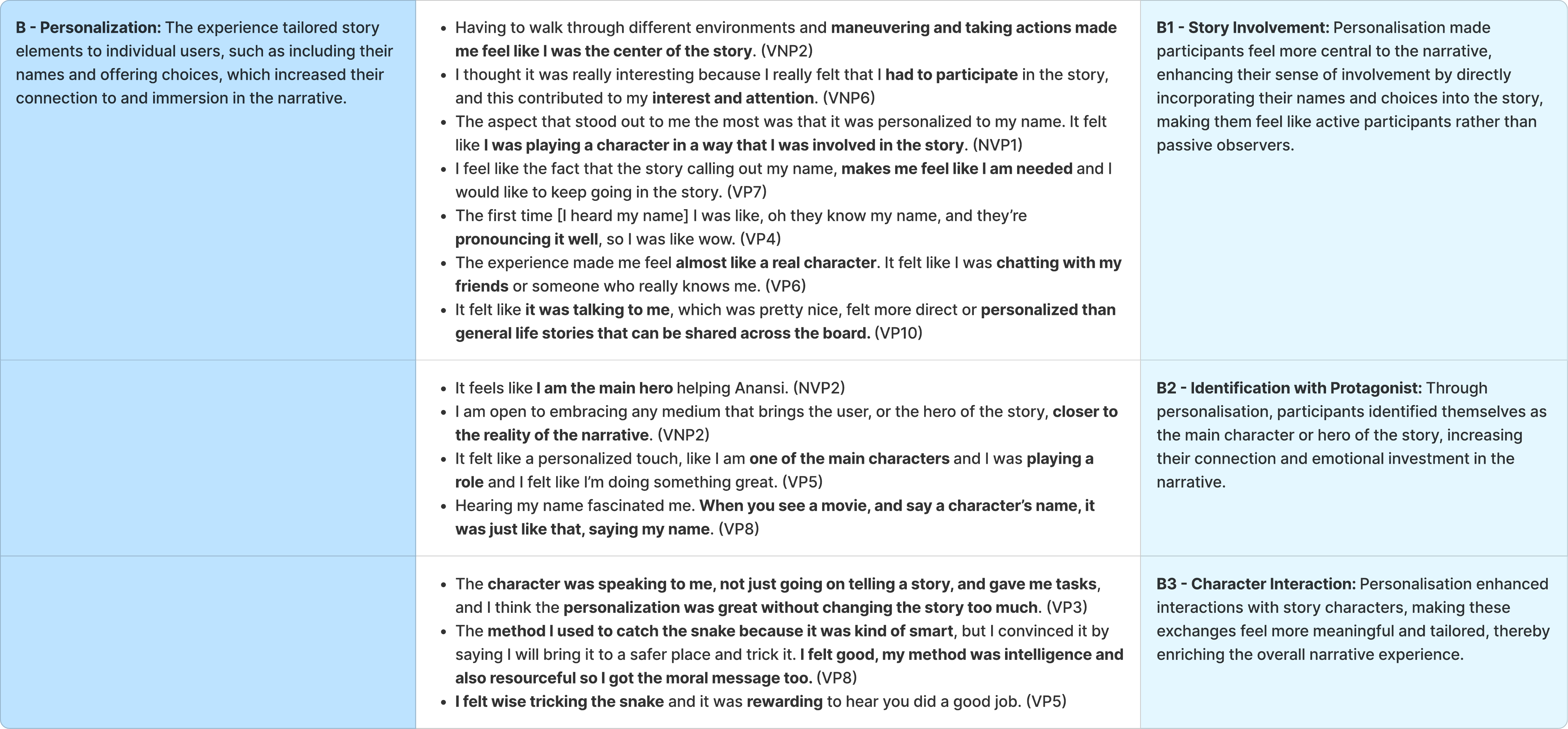}
  \caption{Thematic analysis results: Personalization}
  \label{fig:themes_personalization}
  \Description{This table presenting the theme Personalization, its sub-themes, and representative quotes from participants.}
\end{figure*}

\begin{figure*}[ht]
  \centering
  \includegraphics[width=\linewidth]{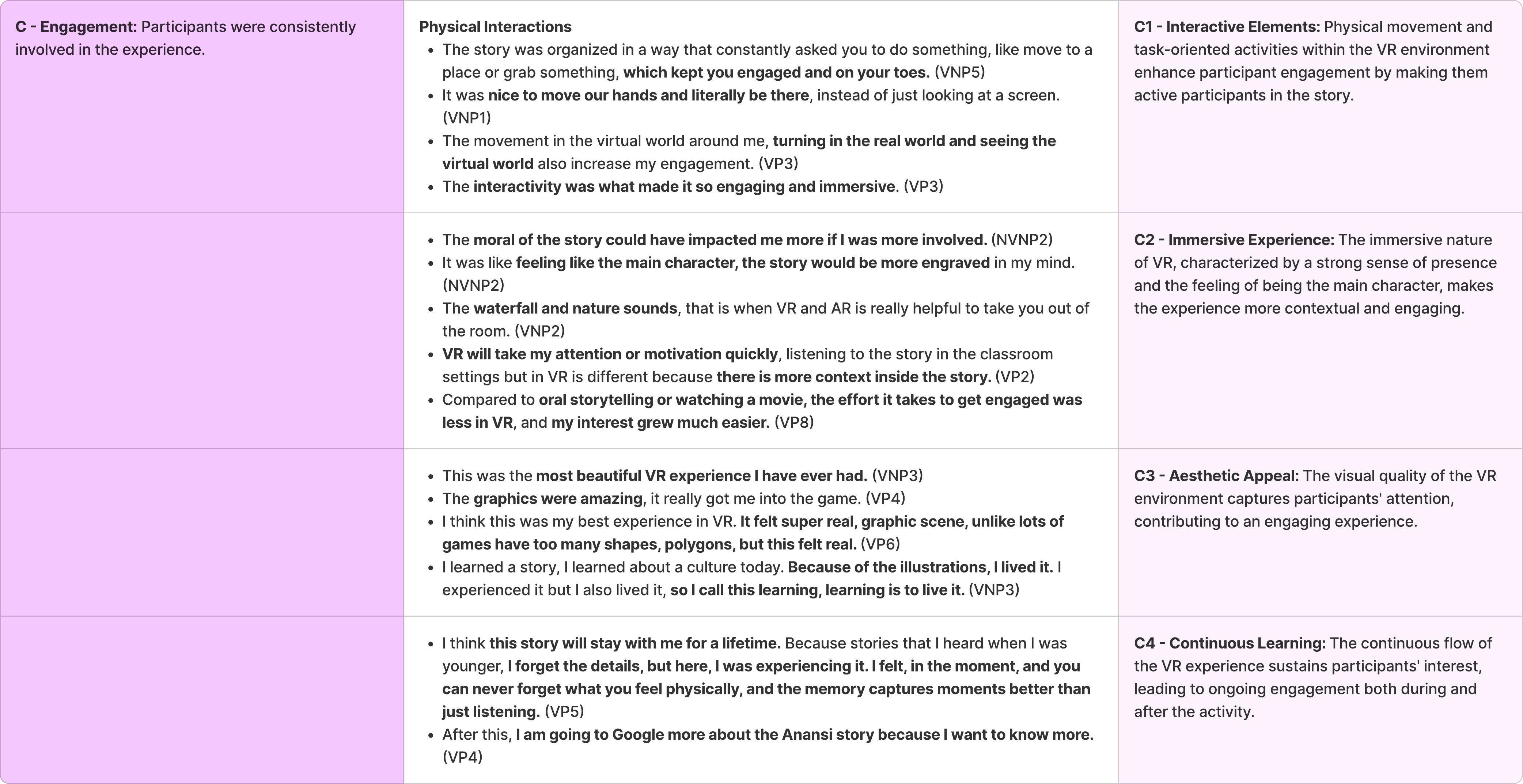}
  \caption{Thematic analysis results: Engagement}
  \label{fig:themes_engagement}
  \Description{This table presenting the theme Engagement, its sub-themes, and representative quotes from participants.}
\end{figure*}

\begin{figure*}[ht]
  \centering
  \includegraphics[width=\linewidth]{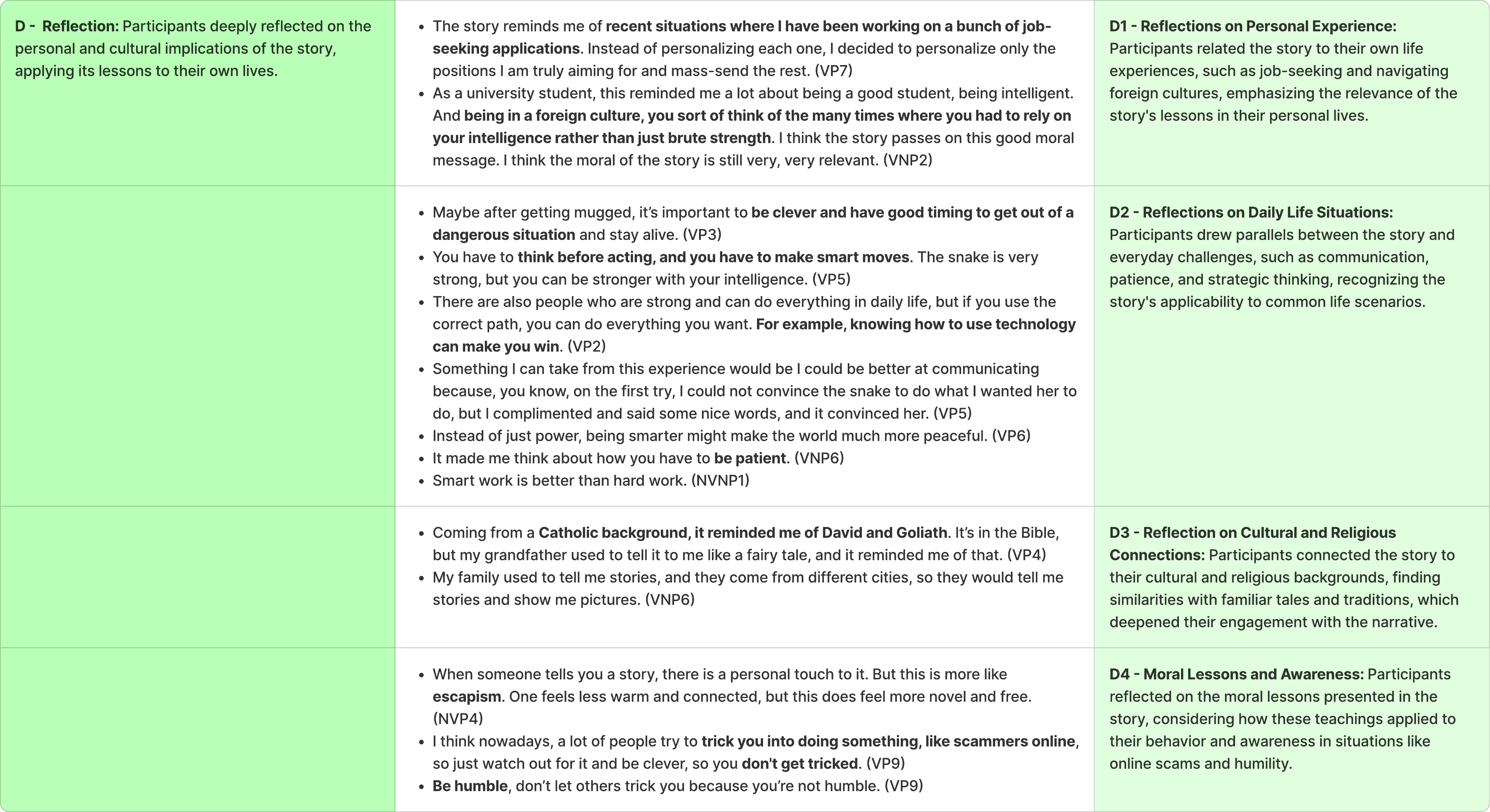}
  \caption{Thematic analysis results: Reflection}
  \label{fig:themes_reflection}
  \Description{This table presenting the theme Reflection, its sub-themes, and representative quotes from participants.}
\end{figure*}

\begin{figure*}[ht]
  \centering
  \includegraphics[width=\linewidth]{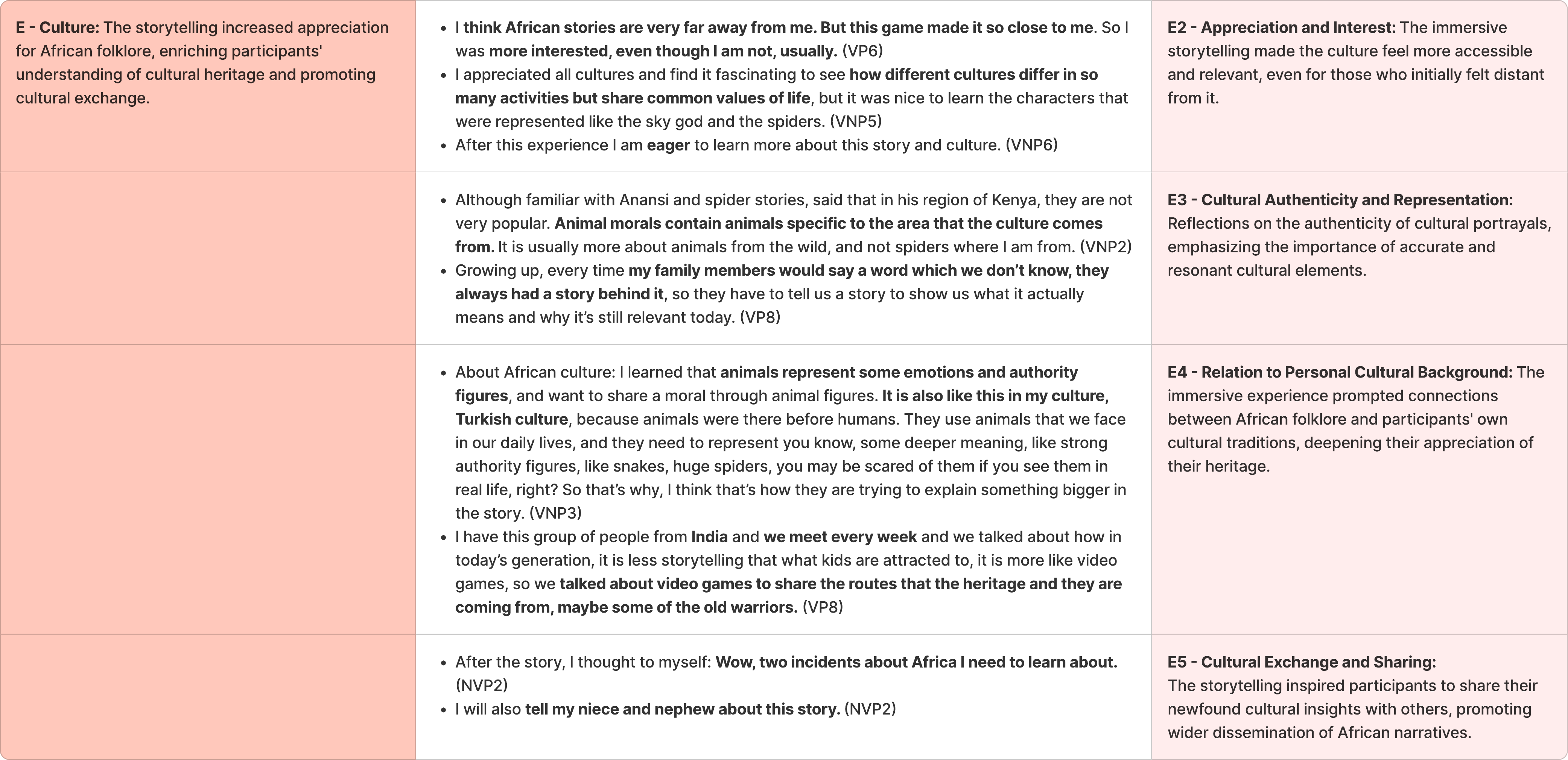}
  \caption{Thematic analysis results: Culture}
  \label{fig:themes_culture}
  \Description{This table presenting the theme Culture, its sub-themes, and representative quotes from participants.}
\end{figure*}

\begin{figure*}[ht]
  \centering
  \includegraphics[width=\linewidth]{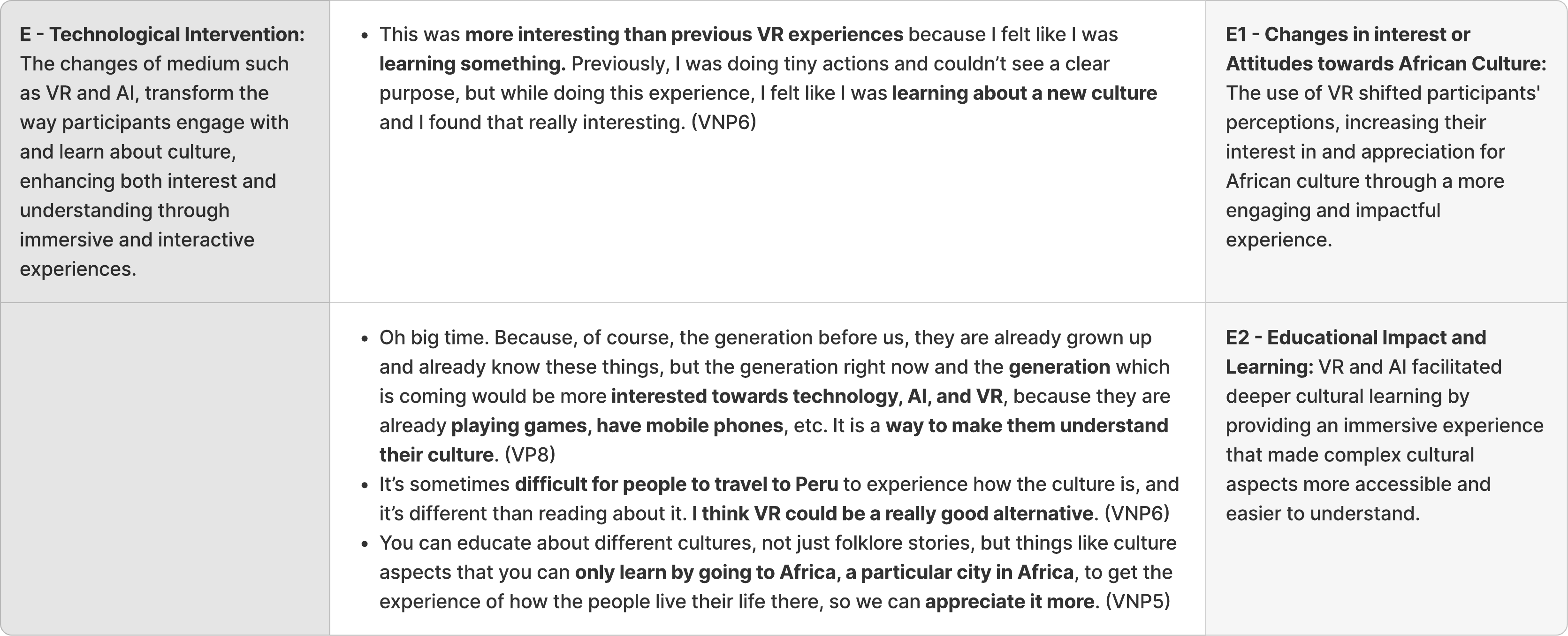}
  \caption{Thematic analysis results: Technological Intervention}
  \label{fig:themes_intervention}
  \Description{This table presenting the theme Technological Intervention, its sub-themes, and representative quotes from participants.}
\end{figure*}

\clearpage
\section{Correlation of UES and Cultural Interests}

\begin{figure*}[ht]
\centering
\includegraphics[width=0.8\linewidth]{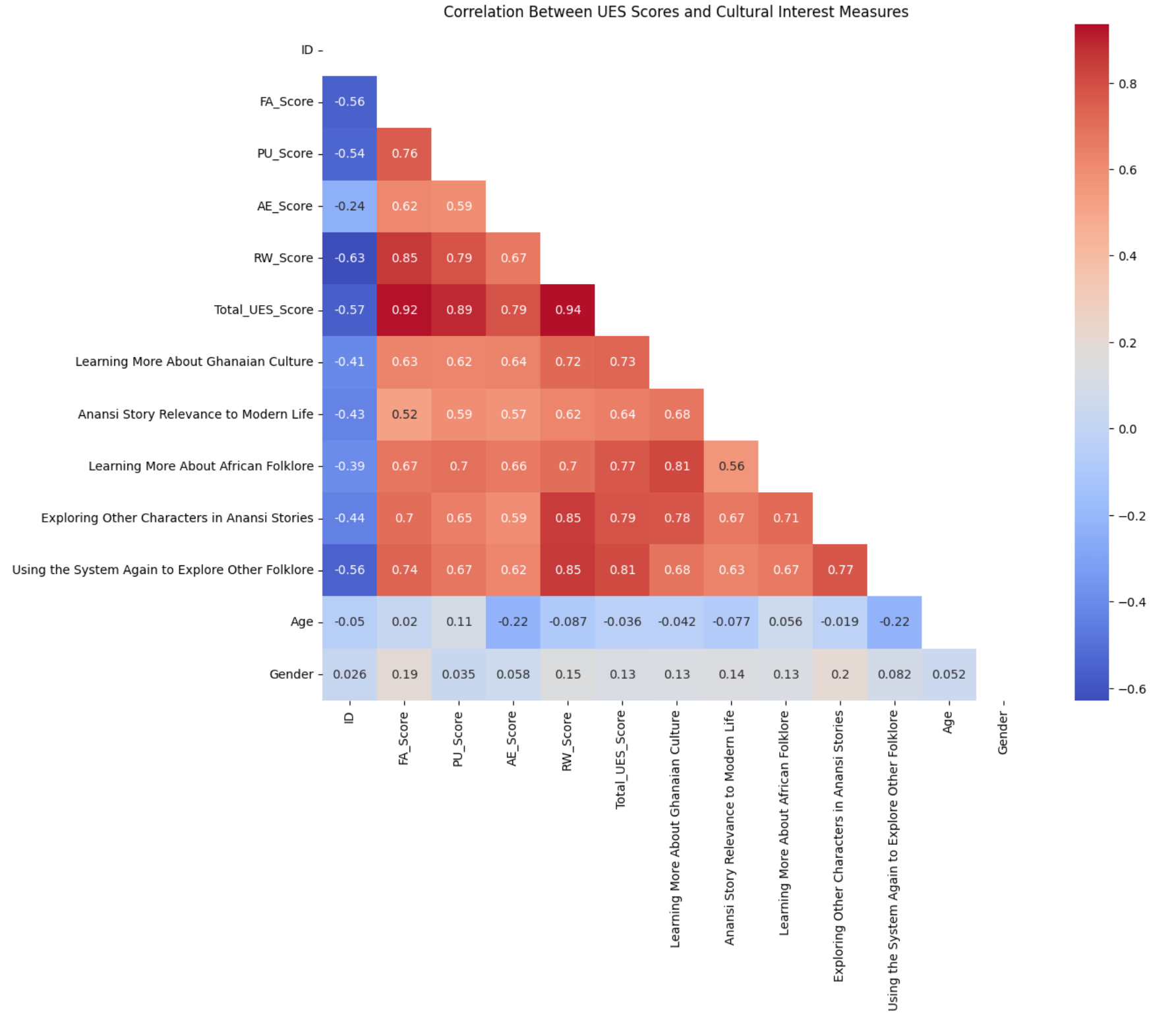}
\caption{Correlation Matrix of User Engagement Scale (UES) Scores and Cultural Interest Measures: This matrix illustrates the relationships between various user engagement metrics and cultural interest variables in the Anansi the Spider VR study. It highlights how different aspects of user engagement correlate with measures of cultural learning and interest, shedding light on the interplay between engagement and cultural learning.}
\Description{This figure illustrates the correlation matrix of User Engagement Scale (UES) Scores and Cultural Interest Measures: This matrix illustrates the relationships between various user engagement metrics and cultural interest variables in the Anansi the Spider VR study. It highlights how different aspects of user engagement correlate with measures of cultural learning and interest, shedding light on the interplay between engagement and cultural learning.}
\label{fig:correlation_metrics}
\end{figure*}

\end{document}